\documentclass[a4paper,11pt]{article}
\pdfoutput=1 % if your are submitting a pdflatex (i.e. if you have
\usepackage{jheppub} % for details on the use of the package, please
\usepackage[T1]{fontenc} % if needed

\usepackage{mathrsfs}
\usepackage{amsmath}
\usepackage{hyperref}

\usepackage{graphicx}% Include figure files
\usepackage{dcolumn}% Align table columns on decimal point
\usepackage{bm}% bold math

\newcommand{\dual}[1]{\overset{\:{}^{^{{{\neg}}}}}{\smash[t]{#1}}} %Elko dual 

\DeclareMathAlphabet{\mathsfit}{T1}{\sfdefault}{\mddefault}{\sldefault}
\SetMathAlphabet{\mathsfit}{bold}{T1}{\sfdefault}{\bfdefault}{\sldefault}

\makeatletter
\newsavebox{\@brx}
\newcommand{\llangle}[1][]{\savebox{\@brx}{\(\m@th{#1\langle}\)}%
  \mathopen{\copy\@brx\mkern2mu\kern-0.9\wd\@brx\usebox{\@brx}}}
\newcommand{\vrt}[1][]{\savebox{\@brx}{\(\m@th{#1\vert}\)}%
  \mathopen{\copy\@brx\mkern2mu\kern-0.9\wd\@brx\usebox{\@brx}}}  
\newcommand{\rrangle}[1][]{\savebox{\@brx}{\(\m@th{#1\rangle}\)}%
  \mathclose{\copy\@brx\mkern2mu\kern-0.9\wd\@brx\usebox{\@brx}}}
\makeatother

\def\p{\mbox{\boldmath$\displaystyle\boldsymbol{p}$}}

\def\bv{\mbox{\boldmath$\displaystyle\boldsymbol{\varphi}$}}

\def\0{\mbox{\boldmath$\displaystyle\boldsymbol{0}$}}
\def\s{\mbox{\boldmath$\displaystyle\boldsymbol{\sigma}$}}

\def\x{\mbox{\boldmath$\displaystyle\boldsymbol{x}$}}
\def\y{\mbox{\boldmath$\displaystyle\boldsymbol{y}$}}

\title{\boldmath Irreducible representations of the Poincar\'e group with reflections and two-fold Wigner degeneracy}

\author[a]{Dharam~Vir~Ahluwalia,}
\author[b]{G.~B.~de~Gracia,}
\author[c]{Julio~M.~Hoff~da~Silva,}
\author[d]{Cheng-Yang Lee,}
\author[e]{B.~M.~Pimentel}
\affiliation[a]{Center for the Studies of the Glass Bead Game, Notting Hill, Victoria 3168, Australia.}
\affiliation[b]{Federal University of ABC, Center of Mathematics,  Santo Andr\'e, 09210-580, Brazil.}
\affiliation[c]{Departamento de Fisica,  Univeridade Estadual Paulista, UNESP, Guaratinguet\'a, SP, Brazil.}
\affiliation[d]{Center for Theoretical Physics, College of Physics, Sichuan University, Chengdu, 610064, China.}
\affiliation[e]{Institute of Theoretical Physics, Sao Paulo State University, 01156-970, S\~ao Paulo,  Brazil. }
\emailAdd{g.gracia@ufabc.edu.br, julio.hoff@unesp.br, cylee@scu.edu.cn, bruto.max@unesp.br}

\abstract{
Not all complete set of spinors can be used as expansion coefficients of a quantum field. In fact, Steven Weinberg established the uniqueness of Dirac spinors for this purpose provided: (a) one paid due attention to the multiplicative phases for each of the spinors, and (b) one paired  these to creation and annihilation operators in a specific manner. This is implicit in his implementation of the rotational symmetry for the spin half quantum field. Among the numerous complete set of spinors that are available to a physicist, Elko occupies a unique status that allows it to enter as expansion coefficients of a quantum field without violating Weinberg's no go theorem. How this paradigm changing claim arises is the primary subject of this communication. Weinberg's no go theorem is evaded by exploiting a uniquely special feature of Elko that allows us to introduce a doubling of the particle-antiparticle degrees of freedom from four to eight. Weinberg had dismissed this degeneracy on the ground that, ``no examples are known of particles that furnish unconventional representations of inversions.'' Here we will find that this degeneracy, once envisioned by Eugene Wigner, in fact gives rise to a quantum field that has all the theoretical properties required of dark matter.
 
 \vspace{0.5cm}
 \begin{quote}\textit{We are deeply saddened to have lost our beloved friend and long-time collaborator Dharam Vir Ahluwalia who has initiated and contributed so much to the theory of mass dimension one fields. His spirit will forever be with us.}
 \end{quote}
}

\begin{document} 
\maketitle
\flushbottom

\section{Introduction}
\label{sec:intro}
In a recent Frontiers Article published in~\cite{Ahluwalia:2022yvk}, we briefly described a fundamentally new particle. For the properties it is endowed with, it is already considered a primary dark matter candidate~\cite{deGracia:2023yit}. It has spin half. The associated quantum field describes particles and antiparticles with eight, compared to four, degrees of the freedom for the Dirac field; that is: it carries a two fold degeneracy once envisioned by Eugene Wigner~\cite{wigner1964unitary}.  Here, we provide full details of the formalism.

The theory of spin half mass dimension one bosons and fermions presents a shift in the established paradigm~\cite{Ahluwalia:2020jkw,Ahluwalia:2022zrm,ahluwalia_2019,Ahluwalia:2022ttu,Ahluwalia:2022yvk}. This work originated from the attempts to understand Majorana spinors~\cite{Ahluwalia:1994uy} which led to the discovery of spin-half fermionic fields of mass dimension one constructed from Elko~\cite{Ahluwalia:2004ab,Ahluwalia:2004sz}.
 They satisfy the spinorial Klein-Gordon but not the Dirac equation and are of mass dimension one. Therefore, they have renormalizable quartic self-interactions. The spin-half bosonic construct seems to contradict the spin-statistics theorem but this concern is unwarranted because, as we will show, due to the non-trivial spinor dual, the mass dimension one bosons and fermions are non-Hermitian. In spite of the non-Hermiticity, the free Hamiltonians have positive-definite real spectra and are bounded from below. Therefore, the spin-statistics theorem which only applies to Hermitian Hamiltonians, is circumvented~\cite{Streater:1989vi,Duck:1997ua}.

These results challenge the long-standing belief that  massive spin-half particles must be described either by a Dirac or a Majorana fermionic field. Yet the `no-go theorem' of Weinberg unambiguously reiterates the prevalent consensus of the physics community, namely the uniqueness of the Dirac and Majorana fermionic fields~\cite{Weinberg:1964cn,Weinberg:1995mt}. Confronted with the no-go theorem, there are two possible outcomes. Either the theory is flawed or there are unexplored features that justify the construct. On the one hand, the theory of Elko and mass dimension one fields have elucidated many different aspects of the Lounesto spinor classification~\cite{daRocha:2005ti,daRocha:2008we,HoffdaSilva:2012uke,daRocha:2013qhu,Bonora:2014dfa,Cavalcanti:2014wia,daRocha:2016bil,HoffdaSilva:2017waf,Fabbri:2017lvu,Meert:2018qzk,Arcodia:2019flm,Rogerio:2020trs,Rogerio:2020cqg,Rogerio:2021ewp,Rogerio:2022tsl}, found applications in cosmology~\cite{Basak:2014qea,Boehmer:2006qq,Boehmer:2007dh,Boehmer:2008ah,Boehmer:2008rz,Boehmer:2009aw,Boehmer:2010ma,HoffdaSilva:2014tth,Pereira:2014wta,S:2014woy,Pereira:2014pqa,Pereira:2016emd,Pereira:2016eez,Pereira:2017efk,Pereira:2017bvq,BuenoRogerio:2017zxf,Pereira:2018xyl,Pereira:2018hir,Pereira:2020ogo,Pereira:2021dkn,Lima:2022vrc}, braneworld models~\cite{Jardim:2014xla,Dantas:2015mfi,Zhou:2017bbj,ZhouZhouXiangNan:2018het,Sorkhi:2018jhy,MoazzenSorkhi:2020fqp} and models of self-interacting dark matter~\cite{Dias:2010aa,Agarwal:2014oaa,Alves:2014kta,Alves:2014qua,Alves:2017joy,Moura:2021rmf}. But on the other hand, our knowledge of the continuous and discrete symmetries of the mass dimension one physical states and the associated quantum fields are incomplete.

Here, we establish the foundations of the theory. We prove that the spin-half mass dimension one fields are irreducible representations of the Poincar\'e group with reflections thus placing them on equal status with the Dirac and Majorana fields. Such representations were first considered by Wigner~\cite{wigner1964unitary} and later studied by Weinberg~\cite[Ch.~2,~App. C]{Weinberg:1995mt}. The associated physical states, are labelled by $p$, $\sigma=\pm\frac{1}{2}$ and an additional Wigner degeneracy index $n=\pm$ so that they are denoted by $|p,\sigma,n\rangle$. With respect to the SM states, they represent a new class of states whose quantum field theory, to the best of our knowledge, have not been explored. To quote Weinberg~\cite[Ch.~2,~App. C]{Weinberg:1995mt}:
\begin{quote}
\textit{No examples are known of particles that furnish unconventional representations of inversions, so these possibilities will not be pursued further here.}
\end{quote}
While this statement is true for the SM, but with about 25\% of the total energy-matter content in the observable universe appearing in the form of dark matter, one should not to be prejudiced against these representations for they may describe particles yet to be discovered. That is, if dark matter are elementary particles beyond the SM, they ought to furnish representations of the Poincar\'{e} group with or without discrete symmetries. We show that the spin-half mass dimension one fields are not only realizations of Wigner degeneracy but are also dark matter candidates.  

The mass dimension one fields are local, Lorentz-covariant and have positive-definite free Hamiltonians. Both the particle and anti-particles have four  degrees of freedom. The doubling of degrees of freedom from two to four is the fundamental point of departure from the Dirac and Majorana constructs that allows us to evade the `no-go' theorem of Weinberg.

%The doubling of degrees of freedom from two to four has important consequences.

The paper is composed as follows. In sec.~\ref{M_to_E}, by identifying the correct degrees of freedom, we construct the complete sets of Elko that are orthonormal, and have Lorentz-invariant spin sums. We construct the quantum field $\lambda$ and its adjoint ${\dual{\lambda}} $ using Elko as expansion coefficients. The fields are local, Lorentz-covariant and have positive-definite free Hamiltonian. By exploiting the freedom in choosing the phases for the dual spinors, the $\lambda$-$\dual{\lambda}$ system can either be bosonic or fermionic. In sec.~\ref{discrete}, we derive the discerete symmetry transformations of the mass dimension one fields and states. In sec.~\ref{Wigner}, the mass dimension one bosons and fermions are shown to be described by states with Wigner degeneracy. Their irreducibility is established from the discrete symmetries. In sec.~\ref{pseudo_h} and app.~\ref{eta}, we take the first step towards formulating a consistent $S$-matrix theory for mass dimension one fields by showing that the term $\dual{\lambda}\lambda$ and the free Lagrangian density to be pseudo Hermitian~\cite{Mostafazadeh:2001jk,Mostafazadeh:2008pw}. In app.~\ref{reducible}, we present a formulation of the theory where the fields constructed from Elko, has eight components and satisfies the Dirac equation. We show that this formulation is reducible so the physics of Elko is described by the four component mass dimension one fields that satisfy the Klein-Gordon equation which furnish the irreducible representation of the Poincar\'e group with reflections with two-fold Wigner degeneracy.

\section{Foundational background}\label{M_to_E}

For this work be widely accessible, we begin with the foundational background. The starting point is Elko. It is an acronym for \textbf{E}igenspinoren des \textbf{L}adungs\textbf{k}onjugations\textbf{o}perators, first introduced in references~\cite{Ahluwalia:2004sz,Ahluwalia:2004ab}. The German to English translation is: Eigenspinors of Charge Conjugation Operator. 

Its existence comes about as follows.\footnote{For notational simplicity, we  confine to spin half. However, the formalism applies to all $(j,0)\oplus(0,j)$ representations.} Taking note of the fact that Wigner time reversal operator  $\Theta$
is defined as 
\begin{equation}
\Theta \pmb{\mathcal{J}}_{R,L} \Theta^{-1} = - \,\pmb{\mathcal{J}}_{R,L}^{*}, 
\end{equation}
where $\pmb{\mathcal{J}}_{R,L}$ are the generators of rotation i the right- and left-handed representations. Reference~\cite{Ahluwalia:1994uy} proved that if $\phi(\p)$ transforms as a left-handed ($\mathcal{L}$) Weyl spinor then  $\Theta\phi^\ast(\p)$ transforms as a right-handed ($\mathcal{R}$) Weyl spinor.  Similarly, if $\phi(\p)$ transforms as a right-handed  Weyl spinor then $\Theta \phi^{*}(\p)$ transforms as a left-handed Weyl spinor. Exploiting the freedom of relative phases between the $\mathcal{L}$-type and $\mathcal{R}$-type Weyl spinors, the following $4$-component $\mathcal{R}\oplus\mathcal{L}$ spinors may be defined as
\begin{align}
\lambda(\p) = \left[\begin{array}{c}
\zeta_\lambda \Theta \phi^{*}(\p)\\
\phi(\p)
\end{array}\right],\quad
\rho(\p) = \left[\begin{array}{c}
\phi(\p)\\
\zeta_\rho \Theta \phi^{*}(\p)
\end{array}\right],
\end{align}
where $\zeta_{\lambda}$ and $\zeta_{\rho}$ are phases to be determined. These become Elko with eigenvalues $\pm 1$, under the action of the charge conjugation operator 
\begin{equation}
\mathcal{C}= \left(
\begin{array}{cc}
\mathbb{O} & i \Theta \\
- i \Theta & \mathbb{O}
\end{array}\label{eq:ChCon}
\right)\mathcal{K},
\end{equation}
where $\mathcal{K}$ stands for the operation of complex conjugation to the right and we take
$\Theta$ in the representation
\begin{align}
\Theta = \left(
\begin{array}{cc}
0 &-1\\
1 & 0
\end{array}
\right).
\end{align}
To avoid any ambiguity and to define the notation, we explicitly note that
\begin{subequations}
\begin{align}
&\mathcal{C}\lambda^{S}(\p)=+\lambda^{S}(\p),\quad
\mathcal{C}\lambda^{A}(\p)=-\lambda^{A}(\p),\label{eq:S}\\
&\mathcal{C}\rho^{S}(\p)=+\rho^{S}(\p),\quad
\mathcal{C}\rho^{A}_{\alpha}(\p)=-\rho^{A}_{\alpha}(\p),  \label{eq:C_psi}
\end{align}
\end{subequations}
where 
\begin{subequations}
\begin{alignat}{2}
\lambda^{S}(\p)&=\lambda(\p)|_{\zeta_{\lambda}=+i},\quad && \lambda^{A}(\p)=\lambda(\p)|_{\zeta_{\lambda}=-i},\label{eq:lSA}\\
\rho^{S}(\p)&=\rho(\p)|_{\zeta_{\rho}=+i},\quad && \rho^{A}(\p)=\rho(\p)|_{\zeta_{\rho}=-i}.\label{eq:rhoSA}
\end{alignat}
\end{subequations}
Elkos $\lambda^{S}$ and $\rho^{S}$ are the self conjugate spinors while $\lambda^{A}$ and $\rho^{A}$ are the anti self conjugate spinors.

To develop the formalism further, we first collect together a few definitions and facts.
We take $\s:=(\sigma_{x},\sigma_{y},\sigma_{z})$ to represent the Pauli matrices.
% in the basis where $\sigma_{z}$ is diagonal
%\begin{equation}
%\sigma_{x}=\left(\begin{matrix}
%0 & 1\\
%1 & 0
%\end{matrix}\right),\quad
%\sigma_{y}=\left(\begin{matrix}
%0 &-i\\
%i & 0
%\end{matrix}\right),\quad
%\sigma_{z}=\left(\begin{matrix}
%1 & 0\\
%0 &-1
%\end{matrix}\right).\label{eq:Pauli}
%\end{equation} 
The generators for 
 the $\mathcal{R}$ and $\mathcal{L}$ irreducible representations of the Lorentz algebra are thus 
 \begin{subequations}
\begin{align}
\mathcal{R}:\qquad\,&{\pmb{\mathcal{J}}}_{R}=\frac{1}{2}\s,\quad{\pmb{\mathcal{K}}}_{R}=-\frac{i}{2}\s,\label{eq:right}\\
\mathcal{L}:\qquad\,&{\pmb{\mathcal{J}}}_{L}=\frac{1}{2}\s,\quad\pmb{\mathcal{K}}_{L}=+\frac{i}{2}\s, \label{eq:left}
\end{align}
\end{subequations}
where $\mathbf{J}$ and $\mathbf{K}_{R,L}$ are the  rotation and boost generators respectively.

Around the unit vector $\hat{\boldsymbol{n}}$, the rotation transformation for the $\mathcal{R}$ as well as $\mathcal{L}$
representation spaces are 
\begin{align}
\exp\left(i{\pmb{\mathcal{J}}}_{R,L}\cdot\boldsymbol{\theta}\right)=\cos\left(\frac{\theta}{2}\right)\mathbb{I}+i(\s\cdot\boldsymbol{\hat{n}})\sin\left(\frac{\theta}{2}\right),\label{eq:rott}
\end{align}
where $\boldsymbol{\theta}=\theta\,\boldsymbol{\hat{n}}$ and $\mathbb{I}$ is the $2\times 2$ identity matrix. The 
$\mathcal{R}$ and 
$\mathcal{L}$ boosts are given by
\begin{subequations}
\begin{align}\mathcal{R}:\quad
&\exp\left(i \pmb{\mathcal{K}}_{R}\cdot\bv\right)=\sqrt{\frac{E+m}{2m}}\left(\mathbb{I}+\frac{\s\cdot\p}{E+m}\right),\label{eq:brt}\\
\mathcal{L}:\quad&\exp\left(i\pmb{\mathcal{K}}_{L}\cdot\bv\right)=\sqrt{\frac{E+m}{2m}}\left(\mathbb{I}-\frac{\s\cdot\p}{E+m}\right),\label{eq:blt}
\end{align}
\end{subequations}
where $\bv:=\varphi\,\boldsymbol{\hat{p}}$, with the rapidity parameter $\varphi$  defined as
\begin{align}
\cosh\varphi=\frac{E_{p}}{m}, \quad \sinh\varphi=\frac{\vert\p\vert}{m}.
\end{align}
The rotation and boost given by~(\ref{eq:rott}) and~(\ref{eq:brt}-\ref{eq:blt}) respectively, specify how the $\mathcal{R}$- and $\mathcal{L}$- Weyl spinors transform. Thus for  the $\mathcal{R}\oplus\mathcal{L}$ representation space the rotation and boost transformations become
\begin{subequations}
\begin{align}
 \mathscr{D}(R(\boldsymbol{\theta})){=}&\exp\left(i{\pmb{\mathcal{J}}}_{R}\cdot\boldsymbol{\theta}\right) \oplus \exp\left(i{\pmb{\mathcal{J}}}_{L}\cdot\boldsymbol{\theta}\right),\\
 \mathscr{D}(L(\p)) {=} &
\exp\left(i\pmb{\mathcal{K}}_{R}\cdot\bv\right)\oplus
\exp\left(i\pmb{\mathcal{K}}_{L}\cdot\bv\right).\label{eq:SpinorBoost}
\end{align} 
\end{subequations}
The associated generators are 
\begin{subequations}
\begin{align}
& \bm{\mathcal{J}}=\left(
\begin{matrix}
\pmb{\mathcal{J}}_{R} & {\mathbb{O}}\\
{\mathbb{O}} & \pmb{\mathcal{J}}_{L}
\end{matrix}
\right)
=\frac12 \left(
\begin{matrix}
\s & \mathbb{O}\\
\mathbb{O} & \s
\end{matrix}
\right),\label{eq:curlyJ}\\
& \bm{\mathcal{K}}=\left(\begin{matrix}
\pmb{\mathcal{K}}_{R} & {\mathbb{O}}\\
{\mathbb{O}} & \pmb{\mathcal{K}}_{L}\end{matrix}\right)=
\frac 12\left(
\begin{matrix}
- i \s & \mathbb{O} \\
\mathbb{O} & i\s
\end{matrix}\right),
\label{eq:generators}
\end{align}
\end{subequations}
where $\mathbb{O}$ is a $2\times 2$ null matrix.

To give a concrete form to Elko, we need $\phi(\p)$. To obtain these, we take $\phi(\0)$ to be the spinors at rest $\p=\0$. We take these to be eigenspinors of $\sigma_z$
\begin{align}
 \sigma_{z}\phi_{\alpha}(\0)=\alpha\phi_{\alpha}(\0), \quad\alpha=\pm,
 \label{eq:alpha}
\end{align}
with a normalization so chosen that in the massless case the rest spinors identically vanish. There are infinitely many choices for such normalisations. We make the choice (that we shall later find to be dimensionally appropriate)
\begin{equation}
 \phi_{+}(\0)=\sqrt{\frac{m}{2}}\left(\begin{matrix}
 1\\
 0
 \end{matrix}\right),\quad
\phi_{-}(\0)=\sqrt{\frac{m}{2}}\left(\begin{matrix}
0\\
1
\end{matrix}\right).\label{eq:phi}
\end{equation}
We do not have to specify whether these transform as $\mathcal{R}$- or $\mathcal{L}$-type Weyl spinors because at rest, the right-handed and left-handed distinction is operationally undefined. At non-zero $\p$, the $\mathcal{R}$-transforming Weyl spinors are
\begin{align}
\phi_\pm(\p) & = \sqrt{\frac{E+m}{2m}}\left(\mathbb{I}+\frac{\s\cdot\p}{E+m}\right) \phi_\pm(\0),
\end{align}
while the $\mathcal{L}$-transforming Weyl spinors are
\begin{align}
\phi_\pm(\p) & = \sqrt{\frac{E+m}{2m}}\left(\mathbb{I}-\frac{\s\cdot\p}{E+m}\right) \phi_\pm(\0).
\end{align}

%The spinors at finite momentum are obtained by applying the boost 
%\begin{align}
%\phi_{\alpha}(\p)&=\exp\left(i\pmb{\mathsfit{K}}_{L}\cdot\bv\right)\phi_{\alpha}(\0). \label{eq:boost}
%\end{align}

%which are eigenspinors of The spinors $\lambda^{S}_{\alpha}(\p)$ and $\lambda^{A}_{\alpha}(\p)$ are the self conjugate and anti self conjugate respectively. They are distinguished by their eigenvalues
%\begin{equation}
%\mathcal{C}\lambda^{S}_{\alpha}(\p)=+\lambda^{S}_{\alpha}(\p),\quad
%\mathcal{C}\lambda^{A}_{\alpha}(\p)=-\lambda^{A}_{\alpha}(\p). \label{eq:C_psi}
%\end{equation}
%In obtaining~(\ref{eq:C_psi}), we have used the fact that $\mathcal{C}$ commutes with the rotation and boost. 
With $\zeta_\lambda$ and $\zeta_\rho$ taking the values given in~(\ref{eq:lSA}-\ref{eq:rhoSA}), following~(\ref{eq:S}-\ref{eq:C_psi}), we thus have $\lambda$-type  Elko
\begin{subequations}
\begin{align}
\lambda^{S}_{+}(\p)&=\left[\begin{matrix}
+i\Theta\phi^{*}_{+}(\p)\\
\phi_{+}(\p)
\end{matrix}\right],\quad
\lambda^{S}_{-}(\p)=\left[\begin{matrix}
+i\Theta\phi^{*}_{-}(\p)\\
\phi_{-}(\p)
\end{matrix}\right],\label{eq:xi}\\
\lambda^{A}_{+}(\p)&=\left[\begin{matrix}
-i\Theta\phi^{*}_{-}(\p)\\
\phi_{-}(\p)
\end{matrix}\right],\quad
\lambda^{A}_{-}(\p)=-\left[\begin{matrix}
-i\Theta\phi^{*}_{+}(\p)\\
\phi_{+}(\p)
\end{matrix}\right].\label{eq:rho}
\end{align}
\end{subequations}
Similarly, the $\rho$-type Elko are
\begin{subequations}
\begin{align}
\rho^{S}_{+}(\p)&=\left[\begin{matrix}
\phi_{+}(\p)\\
-i\Theta\phi^{*}_{+}(\p)
\end{matrix}\right],\quad
\rho^{S}_{-}(\p)=\left[\begin{matrix}
\phi_{-}(\p)\\
-i\Theta\phi^{*}_{-}(\p)\\
\end{matrix}\right],\label{eq:xi2}\\
\rho^{A}_{+}(\p)&=\left[\begin{matrix}
\phi_{-}(\p)\\
+i\Theta\phi^{*}_{-}(\p)
\end{matrix}\right],\quad
\rho^{A}_{-}(\p)=-\left[\begin{matrix}
\phi_{+}(\p)\\
+i\Theta\phi^{*}_{+}(\p)
\end{matrix}\right].\label{eq:rho2}
\end{align}
\end{subequations}

The labels and phases for Elko are chosen to ensure rotational symmetry of the quantum fields to be constructed in sec.~\ref{qf}. This ends the background needed for a crucial observation that now follows.

\subsection{The crucial observation}

%Now comes the crucial observation. 
Because $\mathcal{C}$ is anti-linear, we can construct two additional sets of self conjugate and anti self conjugate Elko  given by 
\begin{subequations}
\begin{align}
\mathcal{C}\left[\epsilon\lambda^{A}_{\alpha}(\p)\right]&=+\left[\epsilon\lambda^{A}_{\alpha}(\p)\right],\label{eq:deg_spinors1}\\
\mathcal{C}\left[\epsilon\lambda^{S}_{\alpha}(\p)\right]&=-\left[\epsilon\lambda^{S}_{\alpha}(\p)\right],
 \label{eq:deg_spinors2}
\end{align}
\end{subequations}
and
\begin{subequations}
\begin{align}
\mathcal{C}\left[\epsilon\rho^{A}_{\alpha}(\p)\right]&=+\left[\epsilon\rho^{A}_{\alpha}(\p)\right],\label{eq:deg_spinors3}\\
\mathcal{C}\left[\epsilon\rho^{S}_{\alpha}(\p)\right]&=-\left[\epsilon\rho^{S}_{\alpha}(\p)\right],
 \label{eq:deg_spinors4}
\end{align}
\end{subequations}
where $\epsilon$ is a purely imaginary phase. Setting $\epsilon= -i$ in~(\ref{eq:deg_spinors1}-\ref{eq:deg_spinors2}), we obtain the extended set of Elko\footnote{We can also obtain the extended set of Elko using~(\ref{eq:deg_spinors3}-\ref{eq:deg_spinors4}). But as these spinors entail the same physics as $\xi$ and $\zeta$, we do not need to consider them.}
\begin{subequations}
\begin{alignat}{2}
\xi(\p,1)&=\lambda^{S}_{+}(\p),\quad &&\xi(\p,2) =\lambda^{S}_{-}(\p),\\
\xi(\p,3)&=\underbrace{-i\lambda^{A}_{+}(\p)}_{\rho^S_+(\bf{p})},\quad &&\xi(\p,4 )=\underbrace{-i\lambda^{A}_{-}(\p)}_{ \rho^S_-(\bf{p})},
\end{alignat}
\end{subequations}
and
\begin{subequations}
\begin{alignat}{2}
\zeta(\p,1)&=\lambda^{A}_{+}(\p),\quad  &&\zeta(\p,2)=\lambda^{A}_{-}(\p),\\
\zeta(\p,3)&=\underbrace{-i\lambda^{S}_{+}(\p)}_{\rho^A_+(\bf{p})},\quad &&\zeta(\p,4)=\underbrace{-i\lambda^{S}_{-}(\p)}_{\rho^A_-(\bf{p})}.
\end{alignat}
\end{subequations}
In the next section, these will be used as expansion coefficients for a spin one half quantum field. This field by virtue of a non-trivial doubling the degrees of freedom results in: (a) evading the Weinberg's no go theorem mentioned in the Abstract, and (b) solving a two decade long problem of not only satisfying  the boost, and translational symmetry, but also the rotational symmetry. This constitutes a major breakthrough not only in the formal theory of quantum fields but it also provides a natural dark matter candidate.

To establish rotational invariance, we shall need the Elko at rest.
These are given by
\begin{subequations}
\begin{alignat}{2}
\xi(\0,1)&=\sqrt{\frac{m}{2}}\left(\begin{matrix}
0 \\
i \\
1 \\
0
\end{matrix}\right),\quad &
\xi(\0,2)&=\sqrt{\frac{m}{2}}\left(\begin{matrix}
-i \\
0 \\
0 \\
1
\end{matrix}\right), \label{eq:rszeta12}\\
\xi(\0,3)&=\sqrt{\frac{m}{2}}\left(\begin{matrix}
1 \\
0 \\
0 \\
-i
\end{matrix}\right),\quad &
\xi(\0,4)&=\sqrt{\frac{m}{2}}\left(\begin{matrix}
0 \\
1 \\
i \\
0
\end{matrix}\right),\label{eq:rszeta34}
\end{alignat}
\end{subequations}
and
\begin{subequations}
\begin{alignat}{2}
\zeta(\0,1)&=\sqrt{\frac{m}{2}}\left(\begin{matrix}
i \\
0 \\
0 \\
1
\end{matrix}\right),\quad &
\zeta(\0,2)&=\sqrt{\frac{m}{2}}\left(\begin{matrix}
0 \\
i \\
-1 \\
0
\end{matrix}\right), \label{eq:rsxi12}\\
\zeta(\0,3)&=\sqrt{\frac{m}{2}}\left(\begin{matrix}
0 \\
1 \\
-i \\
0
\end{matrix}\right),\quad &
\zeta(\0,4)&=\sqrt{\frac{m}{2}}\left(\begin{matrix}
-1 \\
0 \\
0 \\
-i
\end{matrix}\right).\label{eq:rsxi34}
\end{alignat}
\end{subequations}
The $\xi$ and $\zeta$ are obtained from the \textit{Elko at rest}~(\ref{eq:rszeta12}-\ref{eq:rsxi34}) as follows
\begin{align}
\xi(\p,\tau) = \mathscr{D}(L(\p)) \xi(\0,\tau) , \quad
\zeta(\p,\tau) = \mathscr{D}(L(\p)) \zeta(\0,\tau),\quad \tau=1,\cdots,4.
\end{align} 
On using (\ref{eq:SpinorBoost}) in conjunction with~(\ref{eq:brt}-\ref{eq:blt}) yields explicit form of  
 $\mathscr{D}(L(\p))$ 
 \begin{align}
 \mathscr{D}(L(\p)) = \sqrt{\frac{E+m}{2m}}\left(\begin{array}{cc}
 \mathbb{I}+\frac{\boldsymbol{\sigma\cdot p}}{E+m} & \mathbb{O} \\
 \mathbb{O} & \mathbb{I}-\frac{\boldsymbol{\sigma\cdot p}}{E+m}
 \end{array}
 \right).
 \end{align}
Under the Dirac dual, Elko norm identically vanishes~\cite{ahluwalia_2019}. Therefore, we define the Elko dual to be 
\begin{subequations}
\begin{align}
\dual{\xi}(\p,\tau)&=\left[\mathcal{D}\xi(\p,\tau)\right]^{\dag}\gamma^{0},\label{eq:dual1}\\
\dual{\zeta}(\p,\tau)&=s\left[-\mathcal{D}\zeta(\p,\tau)\right]^{\dag}\gamma^{0}\label{eq:dual2},
\end{align}
\end{subequations}
where $\mathcal{D}=m^{-1}\gamma_{\mu}p^{\mu}$ is the Dirac operator, and $s$ is the statistics parameter: $s=+1$ gives fermionic fields, and $s= -1$ yields bosonic fields~\cite{Ahluwalia:2022ttu,Ahluwalia:2022zrm}. Introduction of the new dual is not new in this context, but dramatic effect of the statistics parameter $s$ is entirely unexpected. It affects the orthonormality relations (in the process making the Hamiltonian positive definite) and also the spin sums (affecting the statistics).

%it calculations yield
%\begin{alignat}{2}
%\dual{\xi}(\p,1)&=-i\overline{\xi}(\p,2),\quad &&\dual{\xi}(\p,2)=+i\overline{\xi}(\p,1),\label{eq:duala1}\\
%\dual{\xi}(\p,3)&=+i\overline{\xi}(\p,4),\quad &&\dual{\xi}(\p,4)=-i\overline{\xi}(\p,3),
%\end{alignat}
%and
%\begin{alignat}{2}
%\dual{\zeta}(\p,1)&=-is\overline{\zeta}(\p,2),\quad &&\dual{\zeta}(\p,2)=+is\overline{\zeta}(\p,1),\\
%\dual{\zeta}(\p,3)&=+is\overline{\zeta}(\p,4),\quad &&\dual{\zeta}(\p,4)=-is\overline{\zeta}(\p,3).\label{eq:dualb2}
%\end{alignat}
%where $\overline{\xi}=\xi^{\dag}\gamma^{0}$, $\overline{\zeta}=\zeta^{\dag}\gamma^{0}$. 
Under the Elko dual, the Lorentz invariant orthnormality relations are
\begin{subequations}
\begin{align}
\dual{\xi}(\p,\tau)\xi(\p,\tau')&=m\delta_{\tau\tau'},\label{eq:dxi}\\
\dual{\zeta}(\p,\tau)\zeta(\p,\tau')&=-sm\delta_{\tau\tau'},\label{eq:dzeta}
\end{align}
\end{subequations}
and the spin sums take the form
\begin{subequations}
\begin{align}
\sum_{\tau}\xi(\p,\tau)\dual{\xi}(\p,\tau)&=m\mathbb{I}\label{eq:ss1}, \\
\sum_{\tau}\zeta(\p,\tau)\dual{\zeta}(\p,\tau)&=-sm\mathbb{I}, \label{eq:ss2}
\end{align}
\end{subequations}
where $\tau=1,\cdots,4$, and $\mathbb{I}$ is the $4\times 4$ identity matrix. The doubling of degrees of freedom from two to four is a realization of Wigner degeneracy. Without this doubling of the degrees of freedom the rotational constraints \textit{cannot} be satisfied. 

With the doubling in degrees of freedom, the spin sums computed using the Dirac dual are Lorentz covariant. This raises the question as to which dual is the correct choice for describing the kinematics of Elko. We address this issue in~app.\ref{reducible}. There, we show that the kinematics of the quantum fields constructed using the duals defined in~(\ref{eq:dxi}-\ref{eq:dzeta}) are the correct choice as they furnish an irreducible representation. As for the quantum fields constructed using the Dirac dual, they are reducible.

From the duals, the spinor degrees of freedom, and the spin-sums, it is evident that Elko is fundamentally different from the Dirac spinors. This can be further verified by showing that they do not satisfy the Dirac equation. In the chosen Weyl basis, the $\gamma^{\mu}$ matrices are
\begin{equation}
\gamma^{0}=\left(\begin{matrix}
\mathbb{O} & \mathbb{I} \\
\mathbb{I} & \mathbb{O}
\end{matrix}\right),\,
\gamma^{i}=\left(\begin{matrix}
\mathbb{O} &- \sigma^{i} \\
 \sigma^{i} & \mathbb{O}
\end{matrix}\right),\,
\gamma^{5}=\left(\begin{matrix}
\mathbb{I} & \mathbb{O} \\
\mathbb{O} &-\mathbb{I}
\end{matrix}\right). \label{eq:gamma}
\end{equation}
Direct evaluation yields\footnote{We choose the Minkowski metric to be $\eta_{00}=1$, $\eta_{ij}=-\delta_{ij}$.}
\begin{subequations}
\begin{align}
%\begin{array}{ll}
\gamma^{\mu}p_{\mu}\,\xi(\p,1) = +im \xi(\p,2), & \quad \gamma^{\mu}p_{\mu}\,\xi(\p,2) \label{eq:gp1}
= - im \xi(\p,1),\\
\gamma^{\mu}p_{\mu}\,\xi(\p,3)  = - im \xi(\p,4), &\quad \gamma^{\mu}p_{\mu}\,\xi(\p,4)=+im \xi(\p,3),
%\end{array}
\end{align}
and 
\begin{align}
\gamma^{\mu}p_{\mu}\,\zeta(\p,1) &=- im \zeta(\p,2), \quad \gamma^{\mu}p_{\mu}\,\zeta(\p,2) =  + im \zeta(\p,1),\\
\gamma^{\mu}p_{\mu}\,\zeta(\p,3) & =  +im \zeta(\p,4),\quad \gamma^{\mu}p_{\mu}\,\zeta(\p,4) = - im \zeta(\p,3). \label{eq:gp4}
\end{align}
\end{subequations}
Two successive actions of $\gamma^{\mu}p_{\mu}$ on Elko yield the Klein-Gordon equation
\begin{equation}
\left(p^{\mu}p_{\mu}-m^{2}\right)\xi(\p,\tau)=0,\quad
\left(p^{\mu}p_{\mu}-m^{2}\right)\zeta(\p,\tau)=0. \label{eq:kg}
\end{equation}
Because~(\ref{eq:kg}) is the relativistic dispersion relation, it is not guaranteed to be the equation of motion for Elko. To derive the equation of motion, we need to study the quantum field theory constructed from Elko which is what we will accomplish in sec.~\ref{qf}. By computing the propagator, the canonical commutation/anti-commutation relations, and the free Hamiltonian, we obtain a physically well-defined quantum theory of the spin-half mass dimension one fields. In doing so, we are able to conclude definitively that the Klein-Gordon equation is the correct equation of motion for Elko and its quantum fields. Moreover, we find the physical states to be realizations of the Wigner degeneracy. This inevitably raise the question -- \textit{If the Dirac field describes the SM leptons and quarks, what does the mass dimension one fields describe?} We will attempt to address this question in sec.~\ref{Concl}.

%which is twice as many as the Dirac spinors. In the next section, we show that the particle states and quantum fields associated with Elko are realizations of Wigner degeneracy. 
%The doubling of degrees of freedom makes the spin-sums Lorentz-invariant without $\tau$-deformation.
%In our earlier works~\cite{Ahluwalia:2016rwl}, the $\tau$-deformation was introduced to make the spin-sums Lorentz-invariant but this procedure does not have to be implemented here.
%Repeating the above calculations in the helicity basis, we find the orthonormality relations and the spin-sums to be the same.
%In the process, we are able to obtain the  Lorentz-invariant spin-sums for the self-conjugate and anti-self-conjugate spinors without having to use the $\tau$-deformation. 

\subsection{Spin half fields of mass dimension one}\label{qf}

The quantum field constructed from the extended set of Elko is given by
\begin{align}
\lambda(x)=\int\frac{d^{3}p}{(2\pi)^{3}}\frac{1}{\sqrt{2mE_{p}}}
\sum_{\tau}\left[a(\p,\tau)\xi(\p,\tau)e^{-ip\cdot x}
+b^{\dag}(\p,\tau)\zeta(\p,\tau)e^{ip\cdot x}\right]. \label{eq:F}
\end{align}
The $e^{-i p \cdot x}$ and $e^{i p \cdot x}$ are required by the spacetime translation constraint, see~(5.1.15-5.1.16) in~\cite{Weinberg:1995mt}.

For the field $\lambda$ to respect Poincar\'e space-time symmetries, Elko at rest must satisfy the rotational constraints given in~(5.1.25-5.1.26) of~\cite{Weinberg:1995mt}. They indeed do, something which is impossible without introducing a two-fold Wigner degeneracy. To show this we begin 
with the  rotational constraints 
\begin{subequations}
\begin{align}
\sum_{\tau}\xi_{\ell'}(\0,\tau)\boldsymbol{J}_{\tau\tau'}&=\sum_{\ell}\pmb{\mathcal{J}}_{\ell'\ell}\xi_{\ell}(\0,\tau')\label{eq:rot5}, \\
\sum_{\tau}\zeta_{\ell'}(\0,\tau)\boldsymbol{J}^{*}_{\tau\tau'}&=-\sum_{\ell}\pmb{\mathcal{J}}_{\ell'\ell}\zeta_{\ell}(\0,\tau'),\label{eq:rot6}
\end{align}
\end{subequations}
where $\boldsymbol{J}$ is the rotation generator for the states to be determined. From~(\ref{eq:curlyJ}), the $\bm{\mathcal{J}}$ are known
\begin{align}
{\mathcal{J}}_{x}
=\frac12 \left(
\begin{matrix}
\sigma_{x} & \mathbb{O}\\
\mathbb{O} & \sigma_{x}
\end{matrix}
\right), \quad
{\mathcal{J}}_{y}
=\frac12 \left(
\begin{matrix}
\sigma_{y} & \mathbb{O}\\
\mathbb{O} & \sigma_{y}
\end{matrix}
\right),\quad
{\mathcal{J}}_{z}
=\frac12 \left(
\begin{matrix}
\sigma_{z} & \mathbb{O}\\
\mathbb{O} & \sigma_{z}
\end{matrix}
\right),
\end{align}
and with the enumerated Elko at rest, the right hand side of~(\ref{eq:rot5}-\ref{eq:rot6}) can be fully evaluated by calculating the action of $\mathcal{J}_{i}$, $i=x,y,z$, on the Elko at rest:
\begin{subequations}
\begin{align}
\mathcal{J}_{x}\xi(\0,1)=+\frac{i}{2}\xi(\0,3),\quad%\label{eq:Jx}\\
\mathcal{J}_{x}\xi(\0,2)=-\frac{i}{2}\xi(\0,4),\\
\mathcal{J}_{x}\xi(\0,3)=-\frac{i}{2}\xi(\0,1),\quad
\mathcal{J}_{x}\xi(\0,4)=+\frac{i}{2}\xi(\0,2),
\end{align}
\end{subequations}
\begin{subequations}
\begin{align}
\mathcal{J}_{y}\xi(\0,1)=+\frac{i}{2}\xi(\0,2),\quad
\mathcal{J}_{y}\xi(\0,2)=-\frac{i}{2}\xi(\0,1),\\
\mathcal{J}_{y}\xi(\0,3)=+\frac{i}{2}\xi(\0,4),\quad
\mathcal{J}_{y}\xi(\0,4)=-\frac{i}{2}\xi(\0,3),
\end{align}
\end{subequations}
\begin{subequations}
\begin{align}
\mathcal{J}_{z}\xi(\0,1)=-\frac{i}{2}\xi(\0,4),\quad
\mathcal{J}_{z}\xi(\0,2)=-\frac{i}{2}\xi(\0,3),\\
\mathcal{J}_{z}\xi(\0,3)=+\frac{i}{2}\xi(\0,2),\quad
\mathcal{J}_{z}\xi(\0,4)=+\frac{i}{2}\xi(\0,1).\label{eq:Jz}
\end{align}
\end{subequations}
The actions of $\mathcal{J}_i$, $i=x,y,z$, on anti self conjugate Elko at rest is obtained by replacing $\xi$ with $\zeta$ in the above expressions. With these results at hand, the right hand side of~(\ref{eq:rot5}) for $\mathcal{J}_x$ equals
%\begin{equation}
% \frac{\sqrt{m}}{2}\left(
% \begin{array}{cccc}
%i & 0 &0 &1\\
%0 & -i &1 & 0\\
%0 & 1 & -i & 0\\
%1&0&0&i
%\end{array}\right)
%\end{equation}
\begin{equation}
\left(
\begin{array}{cccc}
 \frac{i \sqrt{m}}{2} & 0 & 0 &
   \frac{\sqrt{m}}{2} \\
 0 & -\frac{i \sqrt{m}}{2} & \frac{\sqrt{m}}{2} &
   0 \\
 0 & \frac{\sqrt{m}}{2} & -\frac{i \sqrt{m}}{2} &
   0 \\
 \frac{\sqrt{m}}{2} & 0 & 0 & \frac{i
   \sqrt{m}}{2} \\
\end{array}
\right),\label{eq:rhs}
\end{equation}
while the left hand side can be written as
%\begin{equation}
%\left(
%\begin{array}{cccc}
% 0 & -i \sqrt{m} & \sqrt{m} & 0 \\
% i \sqrt{m} & 0 & 0 & \sqrt{m} \\
 %\sqrt{m} & 0 & 0 & i \sqrt{m} \\
 %0 & \sqrt{m} & -i \sqrt{m} & 0 \\
%\end{array}
%\right) \times \left(\mbox{a $4\times 4$ matrix representing $J_x$}\right)
%\label{eq:lhs}
%\end{equation}
\begin{equation}
\left(
\begin{array}{cccc}
 0 & -i \sqrt{m} & \sqrt{m} & 0 \\
 i \sqrt{m} & 0 & 0 & \sqrt{m} \\
 \sqrt{m} & 0 & 0 & i \sqrt{m} \\
 0 & \sqrt{m} & -i \sqrt{m} & 0 \\
\end{array}
\right) \times 
\left(\
\begin{array}{c}
\mbox{a $4\times 4$ matrix}\\ 
\mbox{representing $J_x$}
\end{array}
\right).
\label{eq:lhs}
\end{equation}
Equating the two, and solving for $J_x$ yields
\begin{align}
J_x= &
\left(
\begin{array}{cccc}
 0 & -i \sqrt{m} & \sqrt{m} & 0 \\
 i \sqrt{m} & 0 & 0 & \sqrt{m} \\
 \sqrt{m} & 0 & 0 & i \sqrt{m} \\
 0 & \sqrt{m} & -i \sqrt{m} & 0 \\
\end{array}
\right)^{-1} 
\left(
\begin{array}{cccc}
 \frac{i \sqrt{m}}{2} & 0 & 0 &
   \frac{\sqrt{m}}{2} \\
 0 & -\frac{i \sqrt{m}}{2} & \frac{\sqrt{m}}{2} &
   0 \\
 0 & \frac{\sqrt{m}}{2} & -\frac{i \sqrt{m}}{2} &
   0 \\
 \frac{\sqrt{m}}{2} & 0 & 0 & \frac{i
   \sqrt{m}}{2} \\
\end{array}
\right)\nonumber\\
=&\left(
\begin{array}{cccc}
 0 & 0 & -\frac{i}{2} & 0 \\
 0 & 0 & 0 & \frac{i}{2} \\
 \frac{i}{2} & 0 & 0 & 0 \\
 0 & -\frac{i}{2} & 0 & 0 \\
\end{array}
\right).
\end{align}
Repeating the same procedure for the
right hand side of the constraint (\ref{eq:rot5}) for $\mathcal{J}_y$  and $\mathcal{J}_z$ as input, provides $J_y$ and $J_z$
\begin{align}
J_y= \left(
\begin{array}{cccc}
 0 & -\frac{i}{2} & 0 & 0 \\
 \frac{i}{2} & 0 & 0 & 0 \\
 0 & 0 & 0 & -\frac{i}{2} \\
 0 & 0 & \frac{i}{2} & 0 \\
\end{array}
\right),\quad
J_z=\left(
\begin{array}{cccc}
 0 & 0 & 0 & \frac{i}{2} \\
 0 & 0 & \frac{i}{2} & 0 \\
 0 & -\frac{i}{2} & 0 & 0 \\
 -\frac{i}{2} & 0 & 0 & 0 \\
\end{array}
\right).
\end{align}
Or, in a compact form
\begin{align}
J_{x}=\frac{i}{2}\left(\begin{matrix}
\mathbb{O} & -\sigma_{z} \\
\sigma_{z} & \mathbb{O}
\end{matrix}\right),\quad
J_{y}=\frac{1}{2}\left(\begin{matrix}
\sigma_{y} & \mathbb{O} \\
\mathbb{O} & \sigma_{y}
\end{matrix}\right), \quad
J_{z}=\frac{i}{2}\left(\begin{matrix}
\mathbb{O} & \sigma_{x} \\
-\sigma_{x} & \mathbb{O}
\end{matrix}\right). \label{eq:rot_gen1}
\end{align}
The rotational constraint~(\ref{eq:rot6}) is solved by the same set of $\boldsymbol{J}$. As required, these $J_i$, $i=x,y,z$, satisfy the spin half $\text{su}(2)$ algebra with doubly degenerate eigenvalues.

Having provided the new quantum field, we now use the duals defined in~(\ref{eq:dual1}-\ref{eq:dual2}) to introduce
 the adjoint of $\lambda$ 
\begin{align}
\dual{\lambda}(x)=\int\frac{d^{3}p}{(2\pi)^{3}}\frac{1}{\sqrt{2mE_{p}}}
\sum_{\tau}\left[a^{\dag}(\p,\tau)\dual{\xi}(\p,\tau)e^{ip\cdot x}
+b(\p,\tau)\dual{\zeta}(\p,\tau)e^{-ip\cdot x}\right].
\end{align}
The related rotational constraints are
\begin{align}
\sum_{\tau'}\dual{\xi}_{\ell}(\0,\tau')\boldsymbol{J}^{*}_{\tau'\tau}
&=\sum_{\ell'}\boldsymbol{\mathcal{J}}_{\ell'\ell}\dual{\xi}_{\ell'}(\0,\tau),\label{eq:rot11}\\
\sum_{\tau'}\dual{\zeta}_{\ell}(\0,\tau')\boldsymbol{J}_{\tau'\tau}
&= -\sum_{\ell'}\boldsymbol{\mathcal{J}}_{\ell'\ell}\dual{\zeta}_{\ell'}(\0,\tau),\label{eq:rot12}
\end{align}
and the same $\boldsymbol{J}$ satisfy these constraints.
The  $a(\p,\tau)$ and $b^{\dag}(\p,\tau)$ are the annihilation and creation operators for particles and anti-particles respectively satisfying the canonical algebraic relations. The non-vanishing ones are
\begin{align}
\left[a(\p,\tau),a^{\dag}(\p',\tau')\right]_{\pm}&=\left[b(\p,\tau),b^{\dag}(\p',\tau')\right]_{\pm}\nonumber\\
&=(2\pi)^{3}\delta_{\tau\tau'}\delta^{3}(\p-\p').
\end{align}
The top and bottom signs denote anti-commutator and commutator respectively. Using the spin-sums~(\ref{eq:ss1}-\ref{eq:ss2}), we find
\begin{align}
\left[\lambda(t,\x),\dual{\lambda}(t,\y)\right]_{\pm}=&\int\frac{d^{3}p}{(2\pi)^{3}}\frac{1}{2E_{p}}e^{i\boldsymbol{p\cdot(x-y)}}
(1\mp s)\mathbb{I}.
\end{align}
Therefore, locality demands that the statistics parameter $s=\pm1$. When $s=1$, the fields are fermionic, and when $s=-1$, they are bosonic.

%\begin{equation}
%\big\{\lambda(t,\x),\gdual{\lambda}(t,\y)\big\}=O.
%\end{equation}
%The operators satisfy the anti-commutation relations where the non-vanishing ones are

The kinematics is determined by computing the free two-point time-ordered product for $\lambda$ and $\dual{\lambda}$
\begin{align}
\langle\,\,|\mathfrak{T}\big[\lambda(t,\x)\dual{\lambda}(t',\y)\big]|\,\,\rangle=\theta(t-t')\langle\,\,\vert\lambda(t,\x)\dual{\lambda}(t',\y)\vert\,\,\rangle
\mp\theta(t'-t)\langle\,\,\vert\dual{\lambda}(t',\y)\lambda(t,\x)\vert\,\,\rangle,\label{eq:prop}
\end{align}
where $\mathfrak{T}$ is the time-ordering operator. The top and bottom signs in~(\ref{eq:prop}) are for the fermionic and bosonic fields respectively, and $\theta$ is the step-function. A straightforward evaluation using (\ref{eq:ss1}) and (\ref{eq:ss2}) yields
\begin{equation}
\langle\,\,|\mathfrak{T}\lambda(x)\dual{\lambda}(y)|\,\,\rangle=
\frac{i}{(2\pi)^{4}}\int d^{4}q\left[e^{-iq\cdot(x-y)}\frac{\mathbb{I}}{q^{2}-m^{2}+i\epsilon}\right]\nonumber.
\end{equation}
Therefore, the fields $\lambda(x)$ and $\dual{\lambda}(x)$ are of mass dimension one. The propagator is a Green function of the Klein-Gordon operator so the free Lagrangian density is of the form
\begin{equation}
\mathscr{L}_{0}=\left(\partial^{\mu}\dual{\lambda}\partial_{\mu}\lambda-m^{2}\dual{\lambda}\lambda\right).\label{eq:L0}
\end{equation} 
The canonical conjugate momentum for $\lambda$ is $\pi=\partial\dual{\lambda}/\partial t$, so the canonical equal-time relations
\begin{align}
\left[\lambda(t,\x),\lambda(t,\y)\right]_{\pm}&=\left[\pi(t,\x),\pi(t,\y)\right]_{\pm}=\mathbb{O},\\
\left[\lambda(t,\x),\pi(t,\y)\right]_{\pm}&=i\delta^{3}(\x-\y)\mathbb{I}.
\end{align}
To compute the free Hamiltonian, note that the conjugate momentum
 for $\dual{\lambda}$ is $\dual{\pi}=\partial_{t}\lambda$. Therefore,
\begin{align}
H_{0}&=\int d^{3}x\left[\partial_{t}\dual{\lambda}\partial_{t}\lambda-\partial^{i}\dual{\lambda}\partial_{i}\lambda+m^{2}\dual{\lambda}\lambda\right] \nonumber\\
&=\int \frac{d^{3}p}{(2\pi)^3}\,E_{p}\sum_{\tau}\left[a^{\dag}(\p,\tau)a(\p,\tau)-s b(\p,\tau)b^{\dag}(\p,\tau)\right].
\end{align}
The right hand side can be split into two terms. The kinetic part
\begin{align}
H_0{\big\vert}_{\text{kinetic}}= 
\int \frac{d^{3}p}{(2\pi)^3}\,E_{p}\sum_{\tau}
\left[
a^{\dag}(\p,\tau) a(\p,\tau)-s \times 
\begin{cases}  
\mbox{$-\,b^\dagger(\p,\tau) b(\p,\tau)$}\\
\mbox{$+\,b^\dagger(\p,\tau) b(\p,\tau)$}
\end{cases}
\right]
\end{align}
where the upper entry is for fermions, and the bottom part stands
 for bosons. As $s = + 1$ for fermions, and $s=-1$ for bosons the kinetic term reduces to (for fermions as well as bosons)
\begin{align}
H_0{\big\vert}_{\text{kinetic}}= 
\int \frac{d^{3}p}{(2\pi)^3}\,E_{p}\sum_{\tau}
\left[
a^{\dag}(\p,\tau) a(\p,\tau)+
 b^\dagger(\p,\tau) b(\p,\tau)
\right].
\end{align}
It is positive definite. The other part of $H_0$ is the zero-point energy, or the all pervading vacuum energy
\begin{align}
H_0{\big\vert}_{\text{vacuum}}  = 
\int d^{3}p\,E_{p}\sum_{\tau} \delta_{\tau\tau} 
\delta^3(\0),
\end{align}
where the expression for $\delta^3(\0)$ is obtained by observing that 
\begin{equation}
\delta^3(\p)=\frac{1}{(2 \pi)^3}\int d^3x\exp(i\p\cdot\x),
\end{equation} 
formally gives~\cite{Ahluwalia:2004ab}
\begin{equation}
\delta^3(\0) = \frac{1}{(2\pi)^3} \int d^3 x.
\end{equation}
As such
\begin{equation}
H_0{\big\vert}_{\text{vacuum}} = 
-\frac{s}{(2\pi)^3}\int d^{3}p \int d^3x\, (4 \times 2) \left(\frac{E_p}{2}\right).
\end{equation}
In natural units $\hbar=1$ implies $2\pi=h$, $H_0{\big\vert}_{\text{vacuum}}$ therefore represents an energy assignment of $-\frac{s}{2} E_{p}$ for each spinorial degree of freedom (hence the factor of $4$) for the self conjugate and anti self conjugate degrees of freedom (hence the factor of $2$), to each unit-size phase cell $(1/h^3) d^3 p\, d^3 x$ in the sense of statistical mechanics. Its sign is negative for mass dimension one fermions and positive for mass dimension one bosons. 
The two infinities cancel!\footnote{Exactly similar cancellation occurs for the mass dimension three half fields~\cite{Ahluwalia:2020jkw,Ahluwalia:2022zrm}. This makes us suspect if there are spin zero fermions, and vector fermions (massive and massless). If yes, all the infinities associated with quantum fields shall cancel. Then the observed cosmological constant perhaps arises from some quantum corrections.}

\section{Discrete symmetries}\label{discrete}

To establish that the theory of mass dimension one fields is a realization of Wigner degeneracy, 
discrete symmetry transformations for the particle states and fields must be studied. 

Let $|k,\tau,a\rangle$ and $|k,\tau,b\rangle$ be the single particle and anti-particle states at rest with momentum $k^{0}=m$, $\boldsymbol{k}=\0$. In terms of the creation operators, 
\begin{align}
|k,\tau,a\rangle:=&\sqrt{2E_{p}}\,a^{\dag}(\boldsymbol{k},\tau)|\,\,\rangle,\\
|k,\tau,b\rangle:=&\sqrt{2E_{p}}\,b^{\dag}(\boldsymbol{k},\tau)|\,\,\rangle.
\end{align}
Next, let
\begin{equation}
\widehat{\boldsymbol{J}}=(\widehat{J}_{x},\widehat{J}_{y},\widehat{J}_{z}),
\end{equation}
be the rotation generators in the Hilbert space. The action of $\widehat{J}_{z}$ on the states at rest are
\begin{align}
&\widehat{J}_{z}|k,\tau,a\rangle=\sum_{\tau'}(J_{z})_{\tau'\tau}|k,\tau',a\rangle,\\
&\widehat{J}_{z}|k,\tau,b\rangle=\sum_{\tau'}(J_{z})_{\tau'\tau}|k,\tau',b\rangle.
\end{align}
Since $J_{z}$ is block off-diagonal, $|k,\tau,a\rangle$ and $|k,\tau,b\rangle$ are not eigenstates of 
$\widehat{J}_{z}$. To derive the discrete symmetry transformations for $|k,\tau,a\rangle$ and $|k,\tau,b\rangle$, we need to construct the eigenstates of $\widehat{J}_{z}$. Details of the derivations are given in the subsequent sections. The eigenstates are obtained by taking linear combinations of $|k,\tau,a\rangle$ and $|k,\tau,b\rangle$ via
\begin{align}
&\widehat{J}_{z}\sum_{\tau'}S_{\tau'\tau}|k,\tau',a\rangle=\sum_{\tau',\tau''}(S^{-1}J_{z}S)_{\tau'\tau}S_{\tau''\tau'}
|k,\tau'',a\rangle,\\
&\widehat{J}_{z}\sum_{\tau'}S_{\tau'\tau}|k,\tau',b\rangle=\sum_{\tau',\tau''}(S^{-1}J_{z}S)_{\tau'\tau}S_{\tau''\tau'}
|k,\tau'',b\rangle,
\end{align}
where
\begin{equation}
S^{-1}\boldsymbol{J}S=\frac{1}{2}
\left(\begin{matrix}
\s & \mathbb{O}\\
\mathbb{O} & \s
\end{matrix}\right),\quad
S=\frac{1}{\sqrt{2}}\left(\begin{matrix}
\mathbb{I} & \sigma_{y} \\
-\sigma_{y} & \mathbb{I}
\end{matrix}\right). \label{eq:sjs}
\end{equation}
The eigenvalues of $S^{-1}\boldsymbol{J}S$ are thus $\pm\frac{1}{2}$ with a multiplicity of two. These reflect the two-fold degeneracy and shall be represented as $\tau=1,\cdots,4$ in what follows. Thus, the eigenstates of $\widehat{J}_{z}$ are
\begin{align}
&\vrt k,\tau,a\rrangle=\sum_{\tau'}S_{\tau'\tau}|k,\tau',a\rangle,\label{eq:kka}\\
&\vrt k,\tau,b\rrangle=\sum_{\tau'}S_{\tau'\tau}|k,\tau',b\rangle,\label{eq:kkb}
\end{align}
where 
\begin{align}
&\widehat{J}_{z}\vrt k,\tau,a\rrangle=f(\tau)\vrt k,\tau,a\rrangle,
\label{eq:Jzka}\\
&\widehat{J}_{z}\vrt k,\tau,b\rrangle=f(\tau)\vrt k,\tau,b\rrangle,
\label{eq:Jzkb}
\end{align}
with $f(1)=f(3)=\frac{1}{2}$, $f(2)=f(4)=-\frac{1}{2}$. We define the raising and lowering operators to be
\begin{equation}
\widehat{J}_{+}=\widehat{J}_{x}+i\widehat{J}_{y},
\end{equation}
and
\begin{equation}
\widehat{J}_{-}=\widehat{J}_{x}-i\widehat{J}_{y},
\end{equation}
respectively. Acting $\widehat{J}_{\pm}$ on the eigenstates, we get
\begin{align}
\widehat{J}_{-}\vrt k,1,a\rrangle&=\vrt k,2,a\rrangle, \label{eq:Jmk1}\\
\widehat{J}_{-}\vrt k,3,a\rrangle&=\vrt k,4,a\rrangle, \label{eq:Jmk3}
\end{align}
and
\begin{align}
\widehat{J}_{+}\vrt k,2,a\rrangle&=\vrt k,1,a\rrangle, \label{eq:Jpk2}\\
\widehat{J}_{+}\vrt k,4,a\rrangle&=\vrt k,3,a\rrangle. \label{eq:Jpk4}
\end{align}
The same identities apply to the anti-particle states.

\subsection{Charge-conjugation}
 
Let $\mathbf{C}$ be the charge-conjugation operator. By definition, its action is to map the particle states to the anti-particle states and vice versa. Since charge-conjugation is an internal symmetry, its action on the states $\vrt k,\tau,a\rrangle$ and $\vrt k,\tau,b\rrangle$ cannot alter their eigenvalues $\tau$ with respect to $\widehat{J}_{z}$. But due to the two-fold degeneracy as elucidated in~(\ref{eq:Jzka}-\ref{eq:Jzkb}), it may act non-trivially by taking a state, say $\vrt k,1,a\rrangle$ to a linear combination of $\vrt k,1,b\rrangle$ and $\vrt k,3,b\rrangle$. Therefore, the most general charge-conjugation transformation for the states at rest are
\begin{align}
&\mathbf{C}\vrt k,\tau,a\rrangle=\sum_{\tau'}\mathbb{D}_{\tau'\tau}(\mathbf{C})\vrt k,\tau',b\rrangle,\label{eq:ca}\\
&\mathbf{C}\vrt k,\tau,b\rrangle=\sum_{\tau'}\mathbb{\overline{D}}_{\tau'\tau}(\mathbf{C})\vrt k,\tau',a\rrangle.\label{eq:cb}
\end{align}
%where
%\begin{align}
%\mathbb{D}(\mathbf{C})&=\left(\begin{matrix}
%c_{1} & 0 & d_{1} & 0 \\
%0 & c_{2} & 0 & d_{2} \\
%c{3} & 0 & d_{3} & 0 \\
%0 & c_{4} & 0 & d_{4}
%\end{matrix}\right),
%\end{align}
%and
%\begin{align}
%\overline{\mathbb{D}}(\mathbf{C})&=\left(\begin{matrix}
%\bar{c}_{1} & 0 & \bar{d}_{1} & 0 \\
%0 & \bar{c}_{2} & 0 & \bar{d}_{2} \\
%\bar{c}_{3} & 0 & \bar{d}_{3} & 0 \\
%0 & \bar{c}_{4} & 0 & \bar{d}_{4}
%\end{matrix}\right),
%\end{align}
%with $c_{i},d_{i},\bar{c}_{i},\bar{d}_{i}$ being the intrinsic charge-conjugation phases. 
Boosting to arbitrary momentum, we obtain
\begin{align}
&\mathbf{C}\vrt p,\tau,a\rrangle=\sum_{\tau'}\mathbb{D}_{\tau'\tau}(\mathbf{C})\vrt p,\tau',b\rrangle,\\
&\mathbf{C}\vrt p,\tau,b\rrangle=\sum_{\tau'}\mathbb{\overline{D}}_{\tau'\tau}(\mathbf{C})\vrt p,\tau',a\rrangle.
\end{align}
As charge-conjugation is an internal symmetry, the matrices $\mathbb{D}(\mathbf{C})$ and $\overline{\mathbb{D}}(\mathbf{C})$ must commute with the block-diagonal rotation generator~(\ref{eq:sjs}). Solving the commutators, we obtain
\begin{equation}
\mathbb{D}(\mathbf{C})=\left(\begin{matrix}
c_{1}\mathbb{I} & c_{2}\mathbb{I} \\
c_{3}\mathbb{I} & c_{4}\mathbb{I}
\end{matrix}\right),\quad
\overline{\mathbb{D}}(\mathbf{C})=\left(\begin{matrix}
\bar{c}_{1}\mathbb{I} & \bar{c}_{2}\mathbb{I} \\
\bar{c}_{3}\mathbb{I} & \bar{c}_{4}\mathbb{I}
\end{matrix}\right).
\end{equation}
Transform back to the original states,
\begin{align}
&\mathbf{C}\vert p,\tau,a\rangle=\sum_{\tau'}D_{\tau'\tau}(\mathbf{C})\vert p,\tau',b\rangle,\label{eq:ca2}\\
&\mathbf{C}\vert p,\tau,b\rangle=\sum_{\tau'}\overline{D}_{\tau'\tau}(\mathbf{C})\vert p,\tau',a\rangle,\label{eq:cb2}
\end{align}
where
\begin{align}
D(\mathbf{C})&=S\mathbb{D}(\mathbf{C})S^{-1} \nonumber\\
&=\frac{1}{2}\left[\begin{matrix}
(c_{1}+c_{4}) & -i(c_{2}+c_{3}) & (c_{2}-c_{3}) & i(c_{1}-c_{4}) \\
i(c_{2}+c_{3}) & (c_{1}+c_{4}) & -i(c_{1}-c_{4}) & (c_{2}-c_{3}) \\
-(c_{2}-c_{3}) & i(c_{1}-c_{4}) & (c_{1}+c_{4}) & i(c_{2}+c_{3}) \\
-i(c_{1}-c_{4}) & -(c_{2}-c_{3}) & -i(c_{2}+c_{3}) & (c_{1}+c_{4})
\end{matrix}\right],\\
\overline{D}(\mathbf{C})&=D(\mathbf{C})\vert_{c_{i}\rightarrow\bar{c}_{i}}.
\end{align}

We may now derive the charge-conjugation transformation for the mass dimension one field.
%The intrinsic phases can be further constrained by the demand of charge-conjugation symmetry for the mass dimension one fields where $\mathbf{C}\lambda(x)\mathbf{C}^{-1}$ is proportional to $\dual{\lambda}(x)$ up to a constant matrix. To derive the charge-conjugation transformation for $\lambda(x)$, we start by computing $\mathbf{C}\lambda(x)\mathbf{C}^{-1}$. 
Using~(\ref{eq:ca2}-\ref{eq:cb2}), we obtain
\begin{align}
&\mathbf{C}a(\p,\tau)\mathbf{C}^{-1}=\sum_{\tau'}D^{*}_{\tau'\tau}(\mathbf{C})b(\p,\tau'),\label{eq:ca1}\\
&\mathbf{C}b^{\dag}(\p,\tau)\mathbf{C}^{-1}=\sum_{\tau'}\overline{D}_{\tau'\tau}(\mathbf{C})a^{\dag}(\p,\tau'),\label{eq:cb1}
\end{align}
and
\begin{align}
\mathbf{C}\lambda_{\ell}(t,\x)\mathbf{C}^{-1}=\int\frac{d^{3}p}{(2\pi)^{3}}\frac{1}{\sqrt{2mE_{p}}}
\sum_{\tau,\tau'}&\big[D^{*}_{\tau'\tau}(\mathbf{C})b(\p,\tau')\xi_{\ell}(\p,\tau)e^{-ip\cdot x}\nonumber\\
&+\overline{D}_{\tau'\tau}(\mathbf{C})a^{\dag}(\p,\tau')\zeta_{\ell}(\p,\tau)e^{ip\cdot x}\big].\label{eq:C_lambda}
\end{align}
Charge-conjugation is a symmetry because $\mathbf{C}\lambda(x)\mathbf{C}^{-1}$ can be shown to be equal to $\dual{\lambda}(x)$ up to a constant matrix. To prove this, we define $\mathscr{C}=\gamma^{0}\gamma^{2}$ and use the following identities
%Here, we write the adjoint as $\gdual{\lambda}=\lambda^{\#}\gamma^{0}$ where
%\begin{align}
%\lambda^{\#}(t,\x)=&\int\frac{d^{3}p}{(2\pi)^{3}}\frac{1}{\sqrt{2mE_{p}}}\sum_{\tau}\big[e^{-ip\cdot x}\xi^{\#}(\p,\tau)a^{\dag}(\p,\tau)\nonumber\\
%&+e^{ip\cdot x}\zeta^{\#}(\p,\tau)b(\p,\tau)\big],
%\end{align}
%where
%\begin{alignat}{2}
%\xi^{\#}(\p,1)&=-i\xi^{*}(\p,2),\quad &&\xi^{\#}(\p,2)=+i\xi^{*}(\p,1),\\
%\xi^{\#}(\p,3)&=+i\xi^{*}(\p,4),\quad &&\xi^{\#}(\p,4)=-i\xi^{*}(\p,3),
%\end{alignat}
%and
%\begin{alignat}{2}
%\zeta^{\#}(\p,1)&=-is\zeta^{*}(\p,2),\, &&\zeta^{\#}(\p,2)=+is\zeta^{*}(\p,1),\\
%\zeta^{\#}(\p,3)&=+is\zeta^{*}(\p,4),\, &&\zeta^{\#}(\p,4)=-is\zeta^{*}(\p,3).
%\end{alignat}
\begin{align}
\sum_{\ell'}\mathscr{C}_{\ell\ell'}\dual{\xi}_{\ell'}(\p,1)=+\zeta_{\ell}(\p,4),\\
\sum_{\ell'}\mathscr{C}_{\ell\ell'}\dual{\xi}_{\ell'}(\p,2)=-\zeta_{\ell}(\p,3),\\ 
\sum_{\ell'}\mathscr{C}_{\ell\ell'}\dual{\xi}_{\ell'}(\p,3)=+\zeta_{\ell}(\p,2),\\
\sum_{\ell'}\mathscr{C}_{\ell\ell'}\dual{\xi}_{\ell'}(\p,4)=-\zeta_{\ell}(\p,1), 
\end{align}
and
\begin{align}
&\sum_{\ell'}\mathscr{C}_{\ell\ell'}\dual{\zeta}_{\ell'}(\p,1)=-s\xi_{\ell}(\p,4),\\
&\sum_{\ell'}\mathscr{C}_{\ell\ell'}\dual{\zeta}_{\ell'}(\p,2)=+s\xi_{\ell}(\p,3),\\
&\sum_{\ell'}\mathscr{C}_{\ell\ell'}\dual{\zeta}_{\ell'}(\p,3)=-s\xi_{\ell}(\p,2),\\
&\sum_{\ell'}\mathscr{C}_{\ell\ell'}\dual{\zeta}_{\ell'}(\p,4)=+s\xi_{\ell}(\p,1).
\end{align}
Rewrite them in compact forms
\begin{align}
&\sum_{\ell'}\mathscr{C}_{\ell\ell'}\dual{\xi}_{\ell'}(\p,\tau)=f_{c}(\tau)\zeta_{\ell}(\p,\nu_{c}(\tau)),\\
&\sum_{\ell'}\mathscr{C}_{\ell\ell'}\dual{\zeta}_{\ell'}(\p,\tau)=\overline{f}_{c}(\tau)\xi_{\ell}(\p,\overline{\nu}_{c}(\tau)),
\end{align}
we obtain
\begin{align}
\sum_{\ell'}\mathscr{C}_{\ell\ell'}\dual{\lambda}_{\ell'}(x)=\Omega_{C}\int\frac{d^{3}p}{(2\pi)^{3}}\frac{1}{\sqrt{2mE_{p}}}
\sum_{\tau}&\big[a^{\dag}(\p,\tau)f_{c}(\tau)\zeta_{\ell}(\p,\nu_{c}(\tau))e^{ip\cdot x}\nonumber\\
&+b(\p,\tau)\overline{f}_{c}(\tau)\xi_{\ell}(\p,\overline{\nu}_{c}(\tau))e^{-ip\cdot x}\big].\label{eq:g2_lambda}
\end{align}
Multiply~(\ref{eq:g2_lambda}) by an intrinsic charge-conjugation phase $\Omega_{C}$ and equate it to $\mathbf{C}\lambda(x)\mathbf{C}^{-1}$, we obtain
\begin{alignat}{2}
c_{1}&=-c_{4}=is\Omega^{*}_{C},&&\quad c_{2,3}=0,\\
\bar{c}_{1}&=-\bar{c}_{4}=-i\Omega_{C}, &&\quad \bar{c}_{2,3}=0,
\end{alignat}
and hence
\begin{equation}
D(\mathbf{C})=-is\Omega^{*}_{C}\left(\begin{matrix}
\mathbb{O} &\sigma_{y} \\
\sigma_{y} & \mathbb{O}
\end{matrix}\right),\quad
\overline{D}(\mathbf{C})=i\Omega_{C}\left(\begin{matrix}
\mathbb{O} & \sigma_{y} \\
\sigma_{y} & \mathbb{O}
\end{matrix}\right).
\end{equation}
Therefore, the charge-conjugation transformation of the field reads
\begin{equation}
\mathbf{C}\lambda_{\ell}(x)\mathbf{C}^{-1}=\Omega_{C}\sum_{\ell'}\mathscr{C}_{\ell\ell'}\dual{\lambda}_{\ell'}(x).
\end{equation}
For fermions ($s=1)$, the intrinsic charge-conjugation is odd $D^{*}(\mathbf{C})=-\overline{D}(\mathbf{C})$. For bosons ($s=-1)$, it is even $D^{*}(\mathbf{C})=\overline{D}(\mathbf{C})$.

\subsection{Parity}

Let $\mathbf{P}$ be the parity operator. Because $\mathbf{P}\widehat{J}_{z}\mathbf{P}^{-1}=\widehat{J}_{z}$, both $\vrt k,\tau,a\rrangle$ and $\mathbf{P}\vrt k,\tau,a\rrangle$ are eigenstates of $\widehat{J}_{z}$ with the same eigenvalue. Therefore, using~(\ref{eq:Jzka}-\ref{eq:Jzkb}), the most generic parity transformations for the states at rest are given by 
\begin{align}
&\mathbf{P}\vrt k,\tau,a\rrangle=\sum_{\tau'}\mathbb{D}_{\tau'\tau}(\mathbf{P})\vrt k,\tau',a\rrangle,\\
&\mathbf{P}\vrt k,\tau,b\rrangle=\sum_{\tau'}\overline{\mathbb{D}}_{\tau'\tau}(\mathbf{P})\vrt k,\tau',b\rrangle,
\end{align}
where
\begin{equation}
\mathbb{D}(\mathbf{P})=\left(\begin{matrix}
p_{1} & 0 & q_{1} & 0\\
0 & p_{2} & 0 & q_{2}\\
p_{3} & 0 & q_{3} & 0\\
0 & p_{4} & 0 & q_{4} 
\end{matrix}\right),\,
\overline{\mathbb{D}}(\mathbf{P})=\left(\begin{matrix}
\bar{p}_{1} & 0 & \bar{q}_{1} & 0\\
0 & \bar{p}_{2} & 0 & \bar{q}_{2}\\
\bar{p}_{3} & 0 & \bar{q}_{3} & 0\\
0 & \bar{p}_{4} & 0 & \bar{q}_{4} 
\end{matrix}\right),
\end{equation}
with $p_{i},q_{i},\bar{p}_{i},\bar{q}_{i}$ being the intrinsic parity phases. At arbitrary momentum, we find
\begin{align}
&\mathbf{P}\vrt p,\tau,a\rrangle=\sum_{\tau'}\mathbb{D}_{\tau'\tau}(\mathbf{P})\vrt  Pp,\tau',a\rrangle,\\
&\mathbf{P}\vrt p,\tau,b\rrangle=\sum_{\tau'}\overline{\mathbb{D}}_{\tau'\tau}(\mathbf{P})\vrt Pp,{\tau',b}\rrangle,
\end{align}
where $ Pp=(E_{p},-\p)$. Not all intrinsic phases are independent of each other. To determine their relations, let us explicitly write down the parity transformation for each of the particle states. At rest, we have
\begin{align}
\mathbf{P}\vrt k,1,a\rrangle&=p_{1}\vrt k,1,a\rrangle+p_{3}\vrt k,3,a\rrangle,\label{eq:PK1a}\\
\mathbf{P}\vrt k,2,a\rrangle&=p_{2}\vrt k,2,a\rrangle+p_{4}\vrt k,4,a\rrangle, \label{eq:PK2a}\\
\mathbf{P}\vrt k,3,a\rrangle&=q_{3}\vrt k,3,a\rrangle+q_{1}\vrt k,1,a\rrangle,\\
\mathbf{P}\vrt k,4,a\rrangle&=q_{4}\vrt k,4,a\rrangle+q_{2}\vrt k,2,a\rrangle.
\end{align}
%Let $\widehat{J}_{-}=\big(\widehat{J}_{x}-i\widehat{J}_{y}\big)$ be the lowering operator. We note, the states $\vrt k,1,a\rrangle$, $\vrt k,3,a\rangle$ have eigenvalues $+\frac{1}{2}$ and $\vrt k,2,a\rrangle$, $\vrt k,4,a\rrangle$ have eigenvalue $-\frac{1}{2}$ so that
%\begin{equation}
%\widehat{J}_{-}\vrt k,1,a\rrangle=\vrt k,2,a\rrangle,\quad
%\widehat{J}_{-}\vrt k,3,a\rrangle=\vrt k,4,a\rrangle.
%\end{equation}
Applying the lowering operator to~(\ref{eq:PK1a}) and using~(\ref{eq:Jmk1}), we obtain 
\begin{align}
\widehat{J}_{-}\mathbf{P}\vrt k,1,a\rrangle&=\widehat{J}_{-}\left[\,p_{1}\vrt k,1,a\rrangle+p_{3}\vrt k,3,a\rrangle\right]\nonumber\\
&=p_{1}\vrt k,2,a\rrangle+p_{3}\vrt k,4,a\rrangle. \label{eq:JPK1}
\end{align}
Since $[\widehat{J}_{-},\mathbf{P}]=\mathbb{O}$, equation~(\ref{eq:JPK1}) is equivalent to
\begin{align}
\widehat{J}_{-}\mathbf{P}\vrt k,1,a\rrangle&=\mathbf{P}\widehat{J}_{-}\vrt k,1,a\rrangle\nonumber\\
&=\mathbf{P}\vrt k,2,a\rrangle \nonumber\\
&=p_{2}\vrt k,2,a\rrangle+p_{4}\vrt k,4,a\rrangle .\label{eq:PK2b}
\end{align}
Therefore,
\begin{equation}
p_{1}=p_{2},\quad p_{3}=p_{4}.
\end{equation}
Similarly, applying the lowering operator to $\mathbf{P}\vrt k,3,b\rrangle$, we obtain
\begin{equation}
q_{1}=q_{2},\quad q_{3}=q_{4}.
\end{equation}
Repeat the same calculations for the anti-particle states yield
\begin{align}
\bar{p}_{1}=\bar{p}_{2},\quad \bar{p}_{3}=\bar{p}_{4},\quad
\bar{q}_{1}=\bar{q}_{2},\quad \bar{q}_{3}=\bar{q}_{4}.
\end{align}
Transform back to the original states, we obtain
\begin{align}
&\mathbf{P}|p,\tau,a\rangle=\sum_{\tau'}D_{\tau'\tau}|Pp,\tau',a\rangle,\label{eq:pa} \\
&\mathbf{P}|p,\tau,b\rangle=\sum_{\tau'}\overline{D}_{\tau'\tau}|Pp,\tau',b\rangle,\label{eq:pb}
\end{align}
where
\begin{align}
D(\mathbf{P})&=S\mathbb{D}(\mathbf{P})S^{-1} \nonumber\\
&=\frac{1}{2}
\left[\begin{matrix}
(p_{1}+q_{3}) & -i(p_{3}+q_{1}) & -(p_{3}-q_{1}) & i(p_{1}-q_{3}) \\
i(p_{3}+q_{1}) & (p_{1}+q_{3}) & -i(p_{1}-q_{3}) & -(p_{3}-q_{1}) \\
p_{3}-q_{1} & i(p_{1}-q_{3}) & (p_{1}+q_{3}) & i(p_{3}+q_{1}) \\
-i(p_{1}-q_{3}) & (p_{3}-q_{1}) & -i(p_{3}+q_{1}) & (p_{1}+q_{3})
\end{matrix}\right],
\end{align}
and
\begin{align}
\overline{D}(\mathbf{P})&=S\overline{\mathbb{D}}(\mathbf{P})S^{-1} \nonumber \\
&=\left[S\mathbb{D}(\mathbf{P})S^{-1}\right]_{p_{i}\rightarrow\bar{p}_{i},q_{i}\rightarrow\bar{q}_{i}}.
\end{align}

We now derive the parity transformation for the field. Using~(\ref{eq:pa}-\ref{eq:pb}), we obtain
\begin{align}
&\mathbf{P}a(\p,\tau)\mathbf{P}^{-1}=\sum_{\tau'}D^{*}_{\tau'\tau}a(-\p,\tau'),\label{eq:pa1}\\
&\mathbf{P}b^{\dag}(\p,\tau)\mathbf{P}^{-1}=\sum_{\tau'}\overline{D}_{\tau'\tau}b^{\dag}(-\p,\tau'),\label{eq:pb1}
\end{align}
and
\begin{align}
\mathbf{P}\lambda_{\ell}(x)\mathbf{P}^{-1}=&\int\frac{d^{3}p}{(2\pi)^{3}}\frac{1}{\sqrt{2mE_{p}}}
\sum_{\tau,\tau'}
\big[D^{*}_{\tau'\tau}(\mathbf{P})a(-\p,\tau')\xi_{\ell}(\p,\tau)e^{-ip\cdot x}\nonumber\\
&\hspace{3.5cm}+\overline{D}_{\tau'\tau}(\mathbf{P})b^{\dag}(-\p,\tau')\zeta_{\ell}(\p,\tau)e^{ip\cdot x}\big]\nonumber\\
=&\int\frac{d^{3}p}{(2\pi)^{3}}\frac{1}{\sqrt{2mE_{p}}}
\sum_{\tau,\tau'}
\big[D^{*}_{\tau'\tau}(\mathbf{P})a(\p,\tau')\xi_{\ell}(-\p,\tau)e^{-ip\cdot (Px)}\nonumber\\
&\hspace{3.5cm}+\overline{D}_{\tau'\tau}(\mathbf{P})b^{\dag}(\p,\tau')\zeta_{\ell}(-\p,\tau)e^{ip\cdot (Px)}\big].\label{eq:P_lambda}
\end{align}
The parity transformed field $\mathbf{P}\lambda(x)\mathbf{P}^{-1}$ is equal to $\lambda(Px)$ where $Px=(t,-\x)$ up to a constant matrix, making parity a symmetry. To show this, let $\mathscr{P}=\gamma^{0}$. The relevant identities are
\begin{align}
\sum_{\ell'}\mathscr{P}_{\ell\ell'}\xi_{\ell'}(\p,1)&=+i\xi_{\ell}(-\p,2),\label{eq:g01}\\
\sum_{\ell'}\mathscr{P}_{\ell\ell'}\xi_{\ell'}(\p,2)&=-i\xi_{\ell}(-\p,1),\\
\sum_{\ell'}\mathscr{P}_{\ell\ell'}\xi_{\ell'}(\p,3)&=-i\xi_{\ell}(-\p,4),\\
\sum_{\ell'}\mathscr{P}_{\ell\ell'}\xi_{\ell'}(\p,4)&=+i\xi_{\ell}(-\p,3),
\end{align}
and
\begin{align}
\sum_{\ell'}\mathscr{P}_{\ell\ell'}\zeta_{\ell'}(\p,1)&=-i\zeta_{\ell}(-\p,2),\\
\sum_{\ell'}\mathscr{P}_{\ell\ell'}\zeta_{\ell'}(\p,2)&=+i\zeta_{\ell}(-\p,1),\\ 
\sum_{\ell'}\mathscr{P}_{\ell\ell'}\zeta_{\ell'}(\p,3)&=+i\zeta_{\ell}(-\p,4),\\
\sum_{\ell'}\mathscr{P}_{\ell\ell'}\zeta_{\ell'}(\p,4)&=-i\zeta_{\ell}(-\p,3).\label{eq:g04}
\end{align}
Rewrite~(\ref{eq:g01}-\ref{eq:g04}) using the compact notation
\begin{align}
&\sum_{\ell'}\mathscr{P}_{\ell\ell'}\xi_{\ell'}(\p,\tau)=f_{p}(\tau)\xi_{\ell}(-\p,\nu_{p}(\tau)), \\
&\sum_{\ell'}\mathscr{P}_{\ell\ell'}\zeta_{\ell'}(\p,\tau)=\overline{f}_{p}(\tau)\zeta_{\ell}(-\p,\overline{\nu}_{p}(\tau)), 
\end{align}
we get
\begin{align}
\sum_{\ell'}\mathscr{P}_{\ell\ell'}\lambda_{\ell'}(Px)
=\int\frac{d^{3}p}{(2\pi)^{3}}\frac{1}{\sqrt{2mE_{p}}}\
\sum_{\tau}&\big[a(\p,\tau)\tau_{p}(\tau)\xi_{\ell}(-\p,\nu_{p}(\tau))e^{-ip\cdot (Px)}\nonumber\\
&+b^{\dag}(\p,\tau)\overline{\tau}_{p}(\tau)\zeta_{\ell}(-\p,\overline{\nu}_{p}(\tau))e^{ip\cdot(Px)}\big].\label{eq:lambda_P}
\end{align}
Multiply~(\ref{eq:lambda_P}) by an intrinsic parity phase $\Omega_{P}$ and equate it with~(\ref{eq:P_lambda}), we obtain
\begin{align}
& p_{1}=q_{3}=0,\quad p^{*}_{3}=q^{*}_{1}=+\Omega_{P} ,\\
&\overline{p}_{1}=\overline{q}_{3}=0,\quad \overline{p}_{3}=\overline{q}_{1}=+\Omega_{P},
\end{align}
so that
\begin{equation}
D(\mathbf{P})=\Omega^{*}_{P}\left(\begin{matrix}
\sigma_{y} & \mathbb{O} \\
\mathbb{O} & -\sigma_{y}
\end{matrix}\right),\,
\overline{D}(\mathbf{P})=\Omega_{P}\left(\begin{matrix}
\sigma_{y} & \mathbb{O} \\
\mathbb{O} & -\sigma_{y}
\end{matrix}\right).
\end{equation}
Therefore, the parity transformation of the field is given by
\begin{equation}
\mathbf{P}\lambda_{\ell}(x)\mathbf{P}^{-1}=\Omega_{P}\sum_{\ell'}\mathscr{P}_{\ell\ell'}\lambda_{\ell'}(Px).
\end{equation}
Direct evaluations show that $D(\mathbf{P})$ and $\overline{D}(\mathbf{P})$ commute with $\boldsymbol{J}$ so the extended Poincar\'{e} algebra is satisfied. Since $D^{*}(\mathbf{P})=-\overline{D}(\mathbf{P})$, the intrinsic parity is odd.

\subsection{Time-reversal}

Let $\mathbf{T}$ be the time-reversal operator. Because $\mathbf{T}\widehat{J}_{z}\mathbf{T}^{-1}=-\widehat{J}_{z}$, the states $\vrt k,\tau,a\rrangle$ and $\mathbf{T}\vrt k,\tau,a\rangle$ have opposite eigenvalue with respect to $\widehat{J}_{z}$. Therefore, using~(\ref{eq:Jzka}-\ref{eq:Jzkb}), the most generic time-reversal transformations for the states at rest are
\begin{align}
&\mathbf{T}\vrt k,\tau,a\rrangle=\sum_{\tau'}\mathbb{D}_{\tau'\tau}(\mathbf{T})\vrt k,\tau',a\rrangle,\\
&\mathbf{T}\vrt k,\tau,b\rrangle=\sum_{\tau'}\overline{\mathbb{D}}_{\tau'\tau}(\mathbf{T})\vrt k,\tau',b\rrangle,
\end{align}
where
\begin{equation}
\mathbb{D}(\mathbf{T})=\left(\begin{matrix}
0 & t_{1} & 0 & s_{1} \\
t_{2} & 0 & s_{2} & 0 \\
0 & t_{3} & 0 & s_{3} \\
t_{4} & 0 & s_{4} & 0 
\end{matrix}\right),\,
\overline{\mathbb{D}}(\mathbf{T})=\left(\begin{matrix}
0 & \bar{t}_{1} & 0 & \bar{s}_{1} \\
\bar{t}_{2} & 0 & \bar{s}_{2} & 0 \\
0 & \bar{t}_{3} & 0 & \bar{s}_{3} \\
\bar{t}_{4} & 0 & \bar{s}_{4} & 0 
\end{matrix}\right),
\end{equation}
with $s_{i},t_{i},\bar{s}_{i},\bar{t}_{i}$ being the intrinsic time-reversal phases. At arbitrary momentum, we find
\begin{align}
&\mathbf{T}\vrt p,\tau,a\rrangle=\sum_{\tau'}\mathbb{D}_{\tau'\tau}(\mathbf{T})\vrt  Pp,\tau',a\rrangle,\\
&\mathbf{T}\vrt p,\tau,b\rrangle=\sum_{\tau'}\overline{\mathbb{D}}_{\tau'\tau}(\mathbf{T})\vrt Pp,\tau',b\rrangle.
\end{align}
To determine the relations between the phases, we write down the the time-reversal transformations for the particle states in the rest frame
\begin{align}
\mathbf{T}\vrt k,1,a\rrangle &=t_{2}\vrt k,2\rrangle+ t_{4} \vrt k,4,a\rrangle, \label{eq:Tk1a}\\
\mathbf{T}\vrt k,2,a\rrangle &=t_{1}\vrt k,1\rrangle+ t_{3} \vrt k,3,a\rrangle, \\
\mathbf{T}\vrt k,3,a\rrangle &=s_{2}\vrt k,2\rrangle+ s_{4} \vrt k,4,a\rrangle, \\
\mathbf{T}\vrt k,4,a\rrangle &=s_{1}\vrt k,1\rrangle+ s_{3} \vrt k,3,a\rrangle.
\end{align}
Since $\mathbf{T}$ is anti-linear, we have $\widehat{J}_{+}\mathbf{T}=-\mathbf{T}\widehat{J}_{-}$. Acting $\widehat{J}_{+}$ on~(\ref{eq:Tk1a}) and using~(\ref{eq:Jmk1}), we obtain
\begin{align}
\widehat{J}_{+}\mathbf{T}\vrt k,1,a\rrangle&=-\mathbf{T}\widehat{J}_{-}\vrt k,1,a\rrangle\nonumber\\
&=-\mathbf{T}\vrt k,2,a\rrangle\nonumber\\
&=-t_{1}\vrt k,1,a\rrangle-t_{3}\vrt k,3,a\rrangle. \label{eq:Tk1b}
\end{align}
The left-hand side of~(\ref{eq:Tk1b}) can also be written as
\begin{align}
\widehat{J}_{+}\mathbf{T}\vrt k,1,a\rrangle&=\widehat{J}_{+}\left[t_{2}\vrt k,2,a\rrangle+t_{4}\vrt k,4,a\rrangle\right] \nonumber\\
&=t_{2}\vrt k,1,a\rrangle+t_{4}\vrt k,3,a\rrangle. \label{eq:Tk1c}
\end{align}
Equating~(\ref{eq:Tk1b}) with~(\ref{eq:Tk1c}) yields
\begin{equation}
t_{1}=-t_{2},\quad t_{3}=-t_{4}.
\end{equation}
Applying ${J}_{+}$ to $\mathbf{T}\vrt k,3,a\rrangle$, we obtain
\begin{equation}
s_{1}=-s_{2},\quad s_{3}=-s_{4}.
\end{equation}
For the anti-particle states,
\begin{equation}
\bar{t}_{1}=-\bar{t}_{2},\quad \bar{t}_{3}=-\bar{t}_{4},\quad
\bar{s}_{1}=-\bar{s}_{2},\quad \bar{s}_{3}=-\bar{s}_{4}.
\end{equation}
Transform back to the original states yield
\begin{align}
&\mathbf{T}|p,\tau,a\rangle=\sum_{\tau'}D_{\tau'\tau}(\mathbf{T})|Pp,\tau,a\rangle,\label{eq:ta}\\
&\mathbf{T}|p,\tau,b\rangle=\sum_{\tau'}\overline{D}_{\tau'\tau}(\mathbf{T})|Pp,\tau',b\rangle,\label{eq:tb}
\end{align}
where
\begin{align}
D(\mathbf{T})&=S\mathbb{D}(\mathbf{T})(S^{*})^{-1} \nonumber\\
&=\frac{1}{2}\left[\begin{matrix}
-i(s_{1}-t_{3}) & -(s_{3}-t_{1}) & i(s_{3}+t_{1}) & (s_{1}+t_{3}) \\
(s_{3}-t_{1}) & -i(s_{1}-t_{3}) & -(s_{1}+t_{3}) & i(s_{3}+t_{1}) \\
-i(s_{3}+t_{1})& (s_{1}+t_{3}) & -i(s_{1}-t_{3}) & (s_{3}-t_{1}) \\
-(s_{1}+t_{3})& -i(s_{3}+t_{1})&-(s_{3}-t_{1}) & -i(s_{1}-t_{3})
\end{matrix}\right]
\end{align}
and
\begin{align}
\overline{D}(\mathbf{T})=D(\mathbf{T})_{s_{i}\rightarrow\overline{s}_{i},t_{i}\rightarrow\bar{t}_{i}}.
\end{align}

The time-reversal symmetry for the mass dimension one field is verified by relating $\mathbf{T}\lambda(x)\mathbf{T}^{-1}$ with $\lambda(Tx)$ where $Tx=(-t,\x)$. Firstly, we use~(\ref{eq:ta}-\ref{eq:tb}) to obtain
\begin{align}
&\mathbf{T}a(\p,\tau)\mathbf{T}^{-1}=\sum_{\tau'}D^{*}_{\tau'\tau}(\mathbf{T})a(-\p,\tau'),\\
&\mathbf{T}b^{\dag}(\p,\tau)\mathbf{T}^{-1}
=\sum_{\tau'}\overline{D}_{\tau'\tau}(\mathbf{T})b^{\dag}(-\p,\tau'),
\end{align}
so that
\begin{align}
\mathbf{T}\lambda_{\ell}(x)\mathbf{T}^{-1}=&\int\frac{d^{3}p}{(2\pi)^{3}}\frac{1}{\sqrt{2mE_{p}}}
\sum_{\tau,\tau'}\big[
D^{*}_{\tau'\tau}(\mathbf{T})\xi^{*}_{\ell}(\p,\tau)a(-\p,\tau')e^{ip\cdot x}\nonumber\\
&\hspace{3.5cm}+\overline{D}_{\tau'\tau}(\mathbf{T})\zeta^{*}_{\ell}(\p,\tau)b^{\dag}(-\p,\tau')e^{-ip\cdot x}\big]\nonumber\\
=&\int\frac{d^{3}p}{(2\pi)^{3}}\frac{1}{\sqrt{2mE_{p}}}\sum_{\tau,\tau'}\big[
D^{*}_{\tau'\tau}(\mathbf{T})\xi^{*}_{\ell}(-\p,\tau)a(\p,\tau')e^{-ip\cdot(Tx)}\nonumber\\
&\hspace{3.5cm}+\overline{D}_{\tau'\tau}(\mathbf{T})\zeta^{*}_{\ell}(-\p,\tau)b^{\dag}(\p,\tau')e^{ip\cdot(Tx)}\big].\label{eq:T_lambda}
\end{align}
Next, we take $\mathscr{T}=\gamma^{0}\gamma^{2}\gamma^{5}$. The relevant identities are
\begin{align}
\sum_{\ell'}\mathscr{T}_{\ell\ell'}\xi_{\ell'}(\p,1)=+\xi^{*}_{\ell}(-\p,3),\label{eq:gt1} \\
\sum_{\ell'}\mathscr{T}_{\ell\ell'}\xi_{\ell'}(\p,2)=+\xi^{*}_{\ell}(-\p,4),\\
\sum_{\ell'}\mathscr{T}_{\ell\ell'}\xi_{\ell'}(\p,3)=-\xi^{*}_{\ell}(-\p,1),\\
\sum_{\ell'}\mathscr{T}_{\ell\ell'}\xi_{\ell'}(\p,4)=-\xi^{*}_{\ell}(-\p,2),
\end{align}
and
\begin{align}
\sum_{\ell'}\mathscr{T}_{\ell\ell'}\zeta_{\ell'}(\p,1)=-\zeta^{*}_{\ell}(-\p,3), \\
\sum_{\ell'}\mathscr{T}_{\ell\ell'}\zeta_{\ell'}(\p,2)=-\zeta^{*}_{\ell}(-\p,4), \\
\sum_{\ell'}\mathscr{T}_{\ell\ell'}\zeta_{\ell'}(\p,3)=+\zeta^{*}_{\ell}(-\p,1), \\
\sum_{\ell'}\mathscr{T}_{\ell\ell'}\zeta_{\ell'}(\p,4)=+\zeta^{*}_{\ell}(-\p,2).\label{eq:gt4}
\end{align}
Rewrite~(\ref{eq:gt1}-\ref{eq:gt4}) compactly as
\begin{align}
\sum_{\ell'}\mathscr{T}_{\ell\ell'}\xi_{\ell'}(\p,\tau)=f_{t}(\tau)\xi^{*}_{\ell}(-\p,\nu_{t}(\tau)),\\
\sum_{\ell'}\mathscr{T}_{\ell\ell'}\zeta_{\ell'}(\p,\tau)=\overline{f}_{t}(\tau)\zeta^{*}_{\ell}(-\p,\overline{\nu}_{t}(\tau)),
\end{align}
we get 
\begin{align}
\sum_{\ell'}\mathscr{T}_{\ell\ell'}\lambda_{\ell'}(Tx)
=\int\frac{d^{3}p}{(2\pi)^{3}}\frac{1}{\sqrt{2mE_{p}}}
\sum_{\tau}&\big[a(\p,\tau)\tau_{t}(\tau)\xi^{*}_{\ell}(-\p,\nu_{t}(\tau))e^{-ip\cdot(Tx)}\nonumber\\
&+b^{\dag}(\p,\tau)\overline{\tau}_{t}(\tau)\zeta^{*}_{\ell}(-\p,\overline{\nu}_{t}(\tau))e^{ip\cdot(Tx)}\big].\label{eq:lambda_tx}
\end{align}
Multiply~(\ref{eq:lambda_tx}) by an intrinsic time-reversal phase $\Omega_{T}$ and equate it with~(\ref{eq:T_lambda}), we obtain
\begin{align}
& s_{1}=0,\quad s^{*}_{3}=i\Omega_{T}, \\
& \overline{s}_{1}=0,\quad \overline{s}_{3}=i\Omega_{T}.
\end{align}
Therefore,
\begin{equation}
D(\mathbf{T})=\Omega^{*}_{T}\left(\begin{matrix}
\mathbb{O} & \mathbb{I} \\
-\mathbb{I} & \mathbb{O}
\end{matrix}\right),\quad
\overline{D}(\mathbf{T})=\left(
\begin{matrix}
\mathbb{O} & -\mathbb{I} \\
\mathbb{I} & \mathbb{O}
\end{matrix}\right)\Omega_{T},
\end{equation}
and
\begin{equation}
\mathbf{T}\lambda_{\ell}(x)\mathbf{T}^{-1}=\Omega_{T}\sum_{\ell'}\mathscr{T}_{\ell\ell'}\lambda_{\ell'}(Tx).
\end{equation}
Here, $D^{*}(\mathbf{T})=-\overline{D}(\mathbf{T})$ so the intrinsic time-reversal is odd.

\subsection{Algebraic structures}

We can now determine the algebraic structures of $\mathbf{C}$, $\mathbf{P}$, and $\mathbf{T}$. While we cannot fix the intrinsic phases $\Omega_{C,P,T}$, it is possible to obtain useful information on the phases from the products of discrete symmetry operators.\footnote{DVA would like to extend his gratitude to an anonymous referee for the insightful report elucidating the structures of discrete symmetries and Wigner classes.} For charge-conjugation, we find
\begin{align}
&\mathbf{C}^{2}|p,\tau,a\rangle=s|p,\tau,a\rangle,\\
&\mathbf{C}^{2}|p,\tau,b\rangle=s|p,\tau,b\rangle,
\end{align}
with $|\Omega_{C}|^{2}=1$. For parity, we have
\begin{align}
&\mathbf{P}^{2}|p,a\rangle=(\Omega^{*}_{P})^{2}|p,\tau,a\rangle,\\
&\mathbf{P}^{2}|p,b\rangle=\Omega_{P}^{2}|p,\tau,b\rangle.
\end{align}
Since $\mathbf{P}$ is unitary, one may be inclined to choose an appropriate normalization so that $\mathbf{P}^{2}$ is equal to the identity $\mathbb{I}$. However, this seems to be an over simplification and in general, cannot be true for both the particle and anti-particle states. If we take $\mathbf{P}'=(\Omega^{*}_{p})^{-1}\mathbf{P}$, then we would have $\mathbf{P}'^{2}|p,\tau,a\rangle=|p,\tau,a\rangle$ but if the phase is complex, the action on the anti-particle state becomes $\mathbf{P}'^{2}|p,\tau,b\rangle=(\Omega_{P}/\Omega^{*}_{p})^{2}|p,\tau,b\rangle$ which is not equal to unity. For time-reversal,
\begin{align}
&\mathbf{T}^{2}|p,\tau,a\rangle=-|p,\tau,a\rangle, \\
&\mathbf{T}^{2}|p,\tau,b\rangle=-|p,\tau,b\rangle,
\end{align}
with $|\Omega_{T}|^{2}=1$ so $\mathbf{T}^{2}=-1$. 

Additional information on the intrinsic parity phase can be obtained by considering the product of charge-conjugation and parity. Since $\mathbf{CP}$ is unitary, it is related to $\mathbf{PC}$ by
\begin{equation}
\mathbf{CP}=\epsilon\mathbf{PC}, \label{eq:cp}
\end{equation}
where $\epsilon$ is a phase to be determined. Multiply~(\ref{eq:cp}) from the right by $\mathbf{P}$, we obtain
\begin{align}
\mathbf{CP}^{2}&=\epsilon \mathbf{PCP}\nonumber\\
&=\epsilon^{2}\mathbf{P^{2}C}.
\end{align}
Since $\mathbf{P}^{2}$ is proportional to the identity, it commutes with $\mathbf{C}$ thus giving us $\mathbf{CP}^{2}=\epsilon^{2}\mathbf{CP}^{2}$. Therefore, we must have $\epsilon=\pm1$ so $\mathbf{C}$ and $\mathbf{P}$ must either commute or anti-commute. Explicit computations of $\mathbf{CP}$ and $\mathbf{PC}$ for mass dimension one bosons and fermions reveals that the intrinsic parity $\Omega_{P}$ must be either real or imaginary so $\mathbf{P}^{2}=\pm\mathbb{I}$. We find, for both bosons and fermions
\begin{alignat}{2}
&\{\mathbf{C},\mathbf{P}\}=\mathbb{O},&&\quad \Omega_{P}=+\Omega^{*}_{P},\\
&[\mathbf{C},\mathbf{P}]=\mathbb{O},&& \quad\Omega_{P}=-\Omega^{*}_{P}.
\end{alignat}
Acting $(\mathbf{CPT})^{2}$ on the states, we find
\begin{align}
&(\mathbf{CPT})^{2}|p,\tau,a\rangle=s(\Omega_{C}\Omega_{P}\Omega_{T})^{2} |p,\tau,a\rangle,\\
&(\mathbf{CPT})^{2}|p,\tau,b\rangle=s(\Omega^{*}_{C}\Omega^{*}_{P}\Omega^{*}_{P})^{2} |p,\tau,b\rangle.
\end{align}
%where $|\Omega_{C}\Omega_{P}\Omega_{T}|^{2}=1$. Therefore, $(\mathbf{CPT})^{2}$ is equal to $1$ and $-1$ for bosons and fermions respectively.

\section{Irreducibility and Wigner degeneracy}\label{Wigner}

The continuous and discrete symmetry transformations for the particle states are
\begin{equation}
U(R)|p,\tau,a\rangle=\sum_{\tau'}D_{\tau'\tau}(R)|Rp,\tau',a\rangle,
\end{equation}
where $D(R)=\exp(i\boldsymbol{J\cdot\theta})$ and
\begin{align}
&\mathbf{C}|p,\tau,a\rangle=\sum_{\tau'}D_{\tau'\tau}(\mathbf{C})|p,\tau',b\rangle,\\
&\mathbf{P}|p,\tau,a\rangle=\sum_{\tau'}D_{\tau'\tau}(\mathbf{P})|Pp,\tau',a\rangle,\\
&\mathbf{T}|p,\tau,a\rangle=\sum_{\tau'}D_{\tau'\tau}(\mathbf{T})|Pp,\tau',a\rangle,
\end{align}
where $D$ are $4\times4$ matrices. The question arises as to whether these matrices are irreducible. That is, does there exists a unitary basis transformation which renders all the $D$ matrices simultaneously block-diagonal? If such a transformation exist, it would mean that our construct is reducible and the states do not describe elementary particles. We now show explicitly that such a transformation does not exist.

\subsection{Irreducibility}

Let us take transformation on the state $|p,\tau,a\rangle$ to be
\begin{equation}
|p,\tau,a\rangle\rightarrow \sum_{\tau'}Q_{\tau'\tau}|p,\tau',a\rangle, \label{eq:Q}
\end{equation}
where $Q$ is a unitary matrix. Because $\mathbf{C}$, $\mathbf{P}$ are linear and $\mathbf{T}$ is anti-linear, their basis transformations associated with~(\ref{eq:Q}) are
\begin{equation}
\boldsymbol{J}\rightarrow Q^{-1}\boldsymbol{J}Q,
\end{equation}
and
\begin{align}
&D(\mathbf{C})\rightarrow Q^{-1}D(\mathbf{C})Q,\\
&D(\mathbf{P})\rightarrow Q^{-1}D(\mathbf{P})Q,\\
&D(\mathbf{T})\rightarrow Q^{-1}D(\mathbf{T})Q^{*}.
\end{align}
To this end, we are able to make $\boldsymbol{J}$, $D(\mathbf{P})$ and $D(\mathbf{T})$ block-diagonal
by choosing
\begin{equation}
Q^{-1}=\frac{1}{2}\left(\begin{matrix}
\mathbb{I}+\sigma_{y} & \mathbb{I}-\sigma_{y} \\
\mathbb{I}-\sigma_{y} & -\mathbb{I}-\sigma_{y}
\end{matrix}\right).
\end{equation}
The results are
\begin{equation}
Q^{-1}\boldsymbol{J}Q=\frac{1}{2}\left(\begin{matrix}
\s & \mathbb{O} \\
\mathbb{O} & \s
\end{matrix}\right),
\end{equation}
and
\begin{align}
&D'(\mathbf{C})=Q^{-1}D(\mathbf{C})Q=is\Omega^{*}_{C}\left(\begin{matrix}
\mathbb{O} & \mathbb{I} \\
\mathbb{I} & \mathbb{O}
\end{matrix}\right),\\
&D'(\mathbf{P})=Q^{-1}D(\mathbf{P})Q=\Omega^{*}_{P}
\left(\begin{matrix}
\mathbb{I} & \mathbb{O} \\
\mathbb{O} & -\mathbb{I}
\end{matrix}\right),\\
&D'(\mathbf{T})=Q^{-1}D(\mathbf{T})Q^{*}=\Omega^{*}_{T}
\left(\begin{matrix}
\sigma_{y} & \mathbb{O}\\
\mathbb{O} & \sigma_{y}
\end{matrix}\right),
\end{align}
so $D'(\mathbf{C})$ is still block off-diagonal. For the representations to be irreducible, it suffices for us to show that $D'(\mathbf{C})$ and $D'(\mathbf{P})$ cannot be simultaneously made block-diagonal by a unitary transformation. Let 
\begin{equation}
R^{-1}=\left(\begin{matrix}
W & X \\
Y & Z
\end{matrix}\right),
\end{equation}
where $W,X,Y,Z$ are $2\times2$ matrices. Unitarity $RR^{\dag}=\mathbb{I}$ demands
\begin{align}
&WW^{\dag}+XX^{\dag}=YY^{\dag}+ZZ^{\dag}=\mathbb{I},\\
&WY^{\dag}+XZ^{\dag}=\mathbb{O}. \label{eq:wy}
\end{align}
From
\begin{align}
&R^{-1}D'(\mathbf{C})R=\left[\begin{matrix}
(XW^{\dag}+WX^{\dag}) & (XY^{\dag}+WZ^{\dag})\\
(ZW^{\dag}+YX^{\dag}) & (ZY^{\dag}+YZ^{\dag})\label{eq:rdr}
\end{matrix}\right],\\
&R^{-1}D'(\mathbf{P})R=\left[\begin{matrix}
(WW^{\dag}-XX^{\dag}) & (WY^{\dag}-XZ^{\dag}) \\
(YW^{\dag}-ZX^{\dag}) & (YY^{\dag}-ZZ^{\dag})\label{eq:rdr'}
\end{matrix}\right],
\end{align}
and requiring them to be block-diagonal yields
\begin{align}
&XY^{\dag}+WZ^{\dag}=\mathbb{O},\\
&WY^{\dag}-XZ^{\dag}=\mathbb{O}.\label{eq:wy2}
\end{align}
Comparing~(\ref{eq:wy}) with~(\ref{eq:wy2}), it follows that
\begin{equation}
WY^{\dag}=XZ^{\dag}=\mathbb{O}.
\end{equation}
In order for $Q$ to be invertible, we require
\begin{equation}
W\neq\mathbb{O},\quad Z\neq\mathbb{O},\quad X=Y=\mathbb{O},
\end{equation}
or
\begin{equation}
X\neq\mathbb{O},\quad Y\neq\mathbb{O},\quad
W=Z=\mathbb{O}.
\end{equation}
Imposing either constraints on $R^{-1}D'(\mathbf{C})R$, we find that in both cases it cannot be made block diagonal. Therefore, the particle states furnish an irreducible representation of the extended Poincar\'{e} group.

\subsection{Wigner degeneracy}

Thus far, we have represented the two-fold degeneracy by $\tau=1,\cdots 4$. From the group representation perspective, the labelling of states with Wigner degeneracy, which may be more familiar to the readers are $\{\sigma,n\}$ with $\sigma=\pm\frac{1}{2}$ and $n=\pm$ following Weinberg~\cite[Sec.~2, App.~C]{Weinberg:1995mt}. Nevertheless, in most situations, it is more convenient to perform calculations in the basis where $\tau=1,\cdots,4$. Once the calculations are done, one can always rewrite the results in terms of $\{\sigma,n\}$ using the following identifications
\begin{align}
\tau&=1\equiv \{+\textstyle{\frac{1}{2}},+\},\quad \tau=3\equiv \{+\textstyle{\frac{1}{2}},-\},\nonumber \\
\tau&=2\equiv \{-\textstyle{\frac{1}{2}},+\},\quad \tau=4\equiv \{-\textstyle{\frac{1}{2}},-\}, \label{eq:deg_id}
\end{align}
such that $\boldsymbol{J}_{\tau\tau'}=\boldsymbol{J}_{(\sigma,n)(\sigma',n')}$. Similarly, the eigenstates of $\widehat{J}_{z}$ are denoted as
\begin{equation}
\vrt p,\tau,a\rrangle=\vrt p,\sigma,n,a\rrangle,\quad
\vrt p,\tau,b\rrangle=\vrt p,\sigma,n,b\rrangle.
\end{equation}
Therefore, summation over $\tau$ is equivalent to summation over the pairs $\{\sigma,n\}$ following~(\ref{eq:deg_id}).  Using~(\ref{eq:kka}-\ref{eq:kkb}) and~(\ref{eq:deg_id}), we obtain
\begin{align}
&|p,\sigma,n,a\rangle=\sum_{\{\tau',\widetilde{n}'\}}S^{-1}_{(\sigma',n')(\sigma,n)}\vrt p,\sigma',n',a\rrangle,\\
&|p,\sigma,n,b\rangle=\sum_{\{\tau',\widetilde{n}'\}}S^{-1}_{(\sigma',n')(\sigma,n)}\vrt p,\sigma',n',b\rrangle.
\end{align}
The states $|p,\sigma,n,a\rangle$ and $|p,\sigma,n,b\rangle$ are associated with the creation operators $a^{\dag}(\p,\sigma,n)$ and $b^{\dag}(\p,\sigma,n)$ respectively. They satisfy the canonical anti-commutation/commutation relations
\begin{align}
\left[a(\p,\sigma,n),a^{\dag}(\p',\sigma',n')\right]_{\pm}&=\left[b(\p,\sigma,n),b^{\dag}(\p',\sigma',n')\right]_{\pm}\nonumber\\
&=(2\pi)^{3}\delta_{\sigma\sigma'}\delta_{nn'}\delta^{3}(\p-\p').
\end{align}
Making similar identifications with the spinors, the field and its adjoint can then be expanded in the $\{\sigma,n\}$ basis. The spin-sums, locality analysis and the propagator can be readily derived. The discrete symmetries transformations for the states are
\begin{align}
&\mathbf{C}|p,\sigma,n,a\rangle=\sum_{m}(-1)^{1/2-\sigma}E_{mn}(\mathbf{C})|p,-\sigma,m,b\rangle,\\
&\mathbf{C}|p,\sigma,n,b\rangle=\sum_{m}(-1)^{1/2-\sigma}\overline{E}_{mn}(\mathbf{C})|p,-\sigma,m,a\rangle,\\
&\mathbf{P}|p,\sigma,n,a\rangle=\sum_{m}(-1)^{1/2-\sigma}iE_{mn}(\mathbf{P})|Pp,-\sigma,m,a\rangle,\\
&\mathbf{P}|p,\sigma,n,b\rangle=\sum_{m}(-1)^{1/2-\sigma}i\overline{E}_{mn}(\mathbf{P})|Pp,-\sigma,m,b\rangle,\\
&\mathbf{T}|p,\sigma,n,a\rangle=\sum_{m}E_{mn}(\mathbf{T})|Pp,\sigma,m,a\rangle,\\
&\mathbf{T}|p,\sigma,n,b\rangle=\sum_{m}\overline{E}_{mn}(\mathbf{T})|Pp,\sigma,m,b\rangle,
\end{align}
%\begin{align}
%&\mathbf{C}|p,\tau,n,a\rangle=\sum_{m}(-1)^{1/2-\tau}E_{mn}(\mathbf{C})|p,-\tau,m,b\rangle,\\
%&\mathbf{C}|p,\tau,n,b\rangle=\sum_{m}(-1)^{1/2-\tau}\overline{E}_{mn}(\mathbf{C})|p,-\tau,m,a\rangle,\\
%&\mathbf{P}|p,\tau,n,a\rangle=\sum_{m}(-1)^{1/2-\tau}iE_{mn}(\mathbf{P})|Pp,-\tau,m,a\rangle,\\
%&\mathbf{P}|p,\tau,n,b\rangle=\sum_{m}(-1)^{1/2-\tau}i\overline{E}_{mn}(\mathbf{P})|Pp,-\tau,m,b\rangle,\\
%&\mathbf{T}|p,\tau,n,a\rangle=\sum_{m}E_{mn}(\mathbf{T})|Pp,\tau,m,a\rangle,\\
%&\mathbf{T}|p,\tau,n,b\rangle=\sum_{m}\overline{E}_{mn}(\mathbf{T})|Pp,\tau,m,b\rangle,
%\end{align*}
where $E$ and $\overline{E}$ are $2\times2$ matrices given by
\begin{alignat}{2}
E(\mathbf{C})&=s\Omega^{*}_{C}\sigma_{x},\quad &
\overline{E}(\mathbf{C})&=-\Omega_{C}\sigma_{x},\\
E(\mathbf{P})&=\Omega^{*}_{P}\sigma_{z},\quad &
\overline{E}(\mathbf{P})&=\Omega_{P}\sigma_{z},\\
E(\mathbf{T})&=i\Omega^{*}_{T}\sigma_{y},\quad &
\overline{E}(\mathbf{T})&=-i\Omega_{T}\sigma_{y}.
\end{alignat}
It should be noted that we are working in a particular basis where the states at rest $|k,\sigma,n,a\rangle$ and $|k,\sigma,n,b\rangle$ are not eigenstates of $\widehat{J}_{z}$. As a result, the $\sigma$-dependent phases and the mapping from $\sigma$ to $-\sigma$ which are usually associated with time-reversal instead appears in the charge-conjugation and parity transformations.

\section{Pseudo Hermitian quantum field theory} \label{pseudo_h}

We have thus far established that the mass dimension one fields respect Lorentz symmetry, are local and irreducible with two-fold Wigner degeneracy. With all these desirable features, there remains one important issue that we have to address. Because $\dual{\lambda}$ is not equal to $\lambda^{\dag}\gamma^{0}$, the product $\dual{\lambda}\lambda$ while being a Lorentz scalar, is \textit{non-Hermitian}. This is not an issue for the free field theory as the free Hamiltonian is Hermitian and positive-definite.\footnote{Similarly, the three-momentum operators are also Hermitian and positive-definite. Since the free Lagrangian density is a Lorentz scalar, we expect the free Lorentz generators to also be Hermitian.} However, the interacting densities constructed from the free fields $\dual{\lambda}\lambda$ in the interacting picture would be non-Hermitian. Therefore, the interacting theory may not be unitary.

To solve this problem, we now take, what is in our opinion, the first step towards developing a consistent $S$-matrix theory for mass dimension one fields.\footnote{A formalism for computing scattering amplitudes and physical observables utilizing the Elko dual along with the generalized unitarity relation was proposed in~\cite{Ahluwalia:2022ttu}. However, the proposed formalism is incomplete as it has not taken the Wigner degeneracy and pseudo Hermiticity into account.} Here and in app.~\ref{eta}, we will prove that the mass dimension one field is a pseudo Hermitian quantum field theory~\cite{Mostafazadeh:2001jk,Mostafazadeh:2008pw}. 
%This property allows us to define an inner-product that is invariant under time translation generated by non-Hermitian but pseudo-Hermitian Hamiltonians. 
As for the tasks of defining appropriate inner-product, unitary $S$-matrix, and computing physical observables, they have yet to be studied. Nevertheless, we believe that the pseudo Hermitian framework is the correct approach towards accomplishing these tasks. The theory is pseduo Hermitian because there exists a Hermitian $\eta$ operator such that
\begin{equation}
\eta\left[\dual{\lambda}(x)\lambda(x)\right]\eta^{-1}=\left[\dual{\lambda}(x)\lambda(x)\right]^{\dag}.\label{eq:eta_eq}
\end{equation}
We work in the interacting picture so the fields are free. Equation~(\ref{eq:eta_eq}) ensures all interactions constructed in terms of $\dual{\lambda}\lambda$ are pseudo Hermitian. Since the free fields evolve under the free Hamiltonian $H_{0}$, for~(\ref{eq:eta_eq}) to hold at all time, the operator $\eta$ must commute with $H_{0}$. To find $\eta$, we expand the left- and right-hand side of~(\ref{eq:eta_eq}) as
\begin{align}
\eta\left[\dual{\lambda}(x)\lambda(x)\right]\eta^{-1}=&
\int\frac{d^{3}p_{1}}{(2\pi)^{3}(2E_{1})}\int\frac{d^{3}p_{2}}{(2\pi)^{3}(2E_{2})}\nonumber\\
&\times\sum_{\tau\tau'}
\Big\{e^{i(p-p')\cdot x}\dual{\xi}(\p,\tau)\xi(\p',\tau')\eta\left[a^{\dag}(\p,\tau)a(\p',\tau')\right]\eta^{-1}\nonumber\\
&+e^{-i(p-p')\cdot x}\dual{\zeta}(\p,\tau)\zeta(\p',\tau')\eta\left[b(\p,\tau)b^{\dag}(\p',\tau')\right]\eta^{-1}\nonumber\\
&+e^{-i(p+p')\cdot x}\dual{\zeta}(\p,\tau)\xi(\p',\tau')\eta\left[b(\p,\tau)a(\p',\tau')\right]\eta^{-1}\nonumber\\
&+e^{i(p+p')\cdot x}\dual{\xi}(\p,\tau)\zeta(\p',\tau')\eta\left[a^{\dag}(\p,\tau)b^{\dag}(\p',\tau')\right]\eta^{-1}\Big\},
\end{align}
and
\begin{align}
\left[\dual{\lambda}(x)\lambda(x)\right]^{\dag}=&
\int\frac{d^{3}p_{1}}{(2\pi)^{3}(2E_{1})}\int\frac{d^{3}p_{2}}{(2\pi)^{3}(2E_{2})}\nonumber\\
&\times\sum_{\tau\tau'}
\Big\{e^{-i(p-p')\cdot x}\left[\dual{\xi}(\p,\tau)\xi(\p',\tau')\right]^{\dag}a^{\dag}(\p',\tau')a(\p,\tau)\nonumber\\
&+e^{i(p-p')\cdot x}\left[\dual{\zeta}(\p,\tau)\zeta(\p',\tau')\right]^{\dag}b(\p',\tau')b^{\dag}(\p,\tau)\nonumber\\
&+e^{i(p+p')\cdot x}\left[\dual{\zeta}(\p,\tau)\xi(\p',\tau')\right]^{\dag}a^{\dag}(\p',\tau')b^{\dag}(\p,\tau)\nonumber\\
&+e^{-i(p+p')\cdot x}\left[\dual{\xi}(\p,\tau)\zeta(\p',\tau')\right]^{\dag}b(\p',\tau')a(\p,\tau)\Big\}.
\end{align}
Equating them yields
\begin{align}
\eta\left[\sum_{\tau\tau'}\dual{\xi}(\p,\tau)\xi(\p',\tau')a^{\dag}(\p,\tau)a(\p',\tau')\right]\eta^{-1}
&=\sum_{\tau\tau'}\left[\dual{\xi}(\p,\tau)\xi(\p',\tau')\right]^{\dag}a^{\dag}(\p',
\tau')a(\p,\tau),\label{eq:identity1}\\
\eta\left[\sum_{\tau\tau'}\dual{\zeta}(\p,\tau)\xi(\p',\tau')b(\p,\tau)b^{\dag}(\p',\tau')\right]\eta^{-1}
&=\sum_{\tau\tau'}\left[\dual{\zeta}(\p,\tau)\zeta(\p',\tau')\right]^{\dag}b(\p',\tau')b^{\dag}(\p,\tau),\\
\eta\left[\sum_{\tau\tau'}\dual{\zeta}(\p,\tau)\xi(\p',\tau')b(\p,\tau)a(\p',\tau')\right]\eta^{-1}
&=\sum_{\tau\tau'}\left[\dual{\xi}(\p,\tau)\zeta(\p',\tau')\right]^{\dag}b(\p',\tau')a(\p,\tau),\\
\eta\left[\sum_{\tau\tau'}\dual{\xi}(\p,\tau)\zeta(\p',\tau')a^{\dag}(\p,\tau)b^{\dag}(\p',\tau')\right]\eta^{-1}
&=\sum_{\tau\tau'}\left[\dual{\zeta}(\p,\tau)\xi(\p',\tau')\right]^{\dag}a^{\dag}(\p',\tau')b^{\dag}(\p,\tau).\label{eq:identity4}
\end{align}
Next, we rewrite the Elko dual as
\begin{align}
\dual{\xi}(\p,\tau)&=\vartheta(\tau)\xi^{\dag}(\p,\rho(\tau))\gamma^{0},\\
\dual{\zeta}(\p,\tau)&=s\vartheta(\tau)\zeta^{\dag}(\p,\rho(\tau))\gamma^{0},\label{eq:id}
\end{align}
where
\begin{align}
&\vartheta(1)=-i,\quad \vartheta(2)=+i,\label{eq:vart1}\\
&\vartheta(3)=+i,\quad \vartheta(4)=-i,
\end{align}
and
\begin{align}
&\rho(1)=2,\quad \rho(2)=1,\\
&\rho(3)=4,\quad \rho(4)=3.\label{eq:id'}
\end{align}
Substituting~(\ref{eq:id}) into~(\ref{eq:identity1}-\ref{eq:identity4}), we obtain
\begin{align}
\eta\left[a^{\dag}(\p,\tau)a(\p',\tau')\right]\eta^{-1}&=\vartheta(\tau)\label{eq:eta1}\vartheta(\rho(\tau'))a^{\dag}(\p,\rho(\tau))a(\p',\rho(\tau')),\\
\eta\left[b(\p,\tau)b^{\dag}(\p',\tau')\right]\eta^{-1}&=\vartheta(\tau)\vartheta(\rho(\tau'))b(\p,\rho(\tau))b^{\dag}(\p',\rho(\tau')),\\
\eta\left[b(\p,\tau)a(\p',\tau')\right]\eta^{-1}&=\vartheta(\tau)\vartheta(\rho(\tau'))b(\p,\rho(\tau))a(\p',\rho(\tau')),\\
\eta\left[a^{\dag}(\p,\tau)b^{\dag}(\p',\tau')\right]\eta^{-1}&=\vartheta(\tau)\vartheta(\rho(\tau'))a^{\dag}(\p,\rho(\tau))b^{\dag}(\p',\rho(\tau')).\label{eq:eta4}
\end{align}
To keep the notations simple, using the identities~(\ref{eq:vart1}-\ref{eq:id'}), we write $\vartheta(\rho(\tau))=\vartheta^{*}(\tau)$. From~(\ref{eq:eta1}-\ref{eq:eta4}), the actions of $\eta$ on the annihilation and creation operators are given by
\begin{align}
\eta a^{\dag}(\p,\tau)\eta^{-1}&=e^{i\alpha(\tau)}\vartheta(\tau)a^{\dag}(\p,\rho(\tau)),\\
\eta a(\p,\tau)\eta^{-1}&=e^{-i\alpha(\tau)}\vartheta^{*}(\tau)a(\p,\rho(\tau)),
\end{align}
and
\begin{align}
\eta b^{\dag}(\p,\tau)\eta^{-1}&=e^{-i\alpha(\tau)}\vartheta^{*}(\tau)b^{\dag}(\p,\rho(\tau)),\\
\eta b(\p,\tau)\eta^{-1}&=e^{i\alpha(\tau)}\vartheta(\tau)b(\p,\rho(\tau)),
\end{align}
where $\alpha$ is a phase to be determined. By demanding the vacuum to be invariant under $\eta$, its actions on the single particle states are given by
\begin{align}
\eta|\p,\tau,a\rangle&=e^{i\alpha(\tau)}\vartheta(\tau)|\p,\rho(\tau),a\rangle,\label{eq:ea}\\
\eta|\p,\tau,b\rangle&=e^{-i\alpha(\tau)}\vartheta^{*}(\tau)|\p,\rho(\tau),b\rangle,
\end{align}
and
%For the computation of the Dyson series, we also require
\begin{align}
\langle \p,\tau,a|\eta&=\langle\p,\rho(\tau),a|e^{i\alpha(\rho(\tau))}\vartheta^{*}(\tau),\\
\langle \p,\tau,b|\eta&=\langle\p,\rho(\tau),b|e^{-i\alpha(\rho(\tau))}\vartheta(\tau).\label{eq:eb}
\end{align}
Without knowing $e^{i\alpha}$, we cannot determine $\eta$. To show that $\eta$ exists, we take $\alpha$ to be $\tau$-independent and $e^{i\alpha}=-1$. The solution for $\eta$ is then given in app.~\ref{eta}. The actions of $\eta$ on the particle and anti-particle states are then given by
\begin{align}
\eta|\p,\tau,a\rangle&=-\vartheta(\tau)|\p,\rho(\tau),a\rangle,\label{eq:etaa}\\
\eta|\p,\tau,b\rangle&=+\vartheta(\tau)|\p,\rho(\tau),b\rangle,
\end{align}
and
\begin{align}
\langle \p,\tau,a|\eta&=+\langle\p,\rho(\tau),a|\vartheta(\tau),\\
\langle \p,\tau,b|\eta&=-\langle\p,\rho(\tau),b|\vartheta(\tau).\label{eq:etab}
\end{align}
Using~(\ref{eq:etaa}-\ref{eq:etab}), a straightforward calculation shows that $\eta$ commutes with $H_{0}$. Its actions on $\dual{\lambda}$ and $\lambda$ are given by
\begin{align}
\eta\dual{\lambda}(x)\eta^{-1}&=\lambda^{\dag}(x)\gamma^{0},\\
\eta\lambda(x)\eta^{-1}&=\dual{\lambda}^{\dag}(x).
\end{align}
From~(\ref{eq:etaa}-\ref{eq:eb}) and~(\ref{eq:eta_ab}), we find $\eta^{2}=\eta^{\dag}\eta=I$, so $\eta$ is Hermitian.

\section{Conclusions}\label{Concl}

After more than almost three decades studying Majorana spinors followed by Elko and mass dimension one fields, we have now established their foundations. The mass dimension one fields and their particle states furnish the unitary irreducible representations of the extended Poincar\'{e} group. 
%The fields $\lambda$ and $\dual{\lambda}$ are Lorentz-covariant and $\dual{\lambda}\lambda$ is a Lorentz scalar.

Comparing to the earlier formulations of the theory, there are important similarities as well as differences. The orthonormal norms and spin-sums of Elko are identical to the ones given under $\tau$-deformation~\cite{Ahluwalia:2022ttu}. The difference, which is one of the main results of this paper, is the doubling of degrees of freedom from two to four for the particles as well as the anti-particles. From the discrete symmetry analysis, this is precisely the degeneracy induced by the inclusion of non-trivial discrete symmetries studied by Wigner~\cite{wigner1964unitary} and later expanded by Weinberg~\cite[Sec.~2~C]{Weinberg:1995mt}. Upon identifying the correct degrees of freedom, it is a straightforward to verify Lorentz covariance for the mass dimension one fields. %This directly addresses the issues raised in~\cite{Gillard:2010nr,Aguirre:2021whp}

%These results are consistent and extend the formalism developed Weinberg.

There remains much to be done. In the spin-half $\mathcal{R}\oplus\mathcal{L}$ representation, the mass dimension one fields constructed from Elko may only be one of the many non-trivial solutions. If more solutions exists, we believe that they can be systematically classified by their discrete symmetries. More generally, one should study quantum field theory with Wigner degeneracy for higher-spin representations of the extended Poincar\'{e} group. In the case of Elko and the associated mass dimension one fields, the extension to higher-spin should be straightforward following~\cite{Lee:2012td}.

Phenomenologically, we believe that the mass dimension one bosons and fermions are natural dark matter candidates. Due to the mismatch in mass dimensions with the SM fermions (for mass dimension one bosons, there is also a mismatch in statistics), they cannot form doublets or multiplets with the SM fermions. As singlets, mass dimension one particles can only interact with the SM fermions via gravity, Higgs or dark gauge fields thus making them natural candidates as self-interacting dark matter. On the other hand, if there are doublets or multiplets comprised of mass dimension one fields, they may interact with the SM sector through the SM gauge fields. We leave these important phenomenological studies for future investigations. 

There are two interactions that distinguish the spin-half mass dimension one fields from their Dirac counterparts. Due to their mass dimensions, the mass dimension one fields are naturally endowed with a renormalizable quartic self-interaction $g(\dual{\lambda}\lambda)^{2}$. There is also quadratic couplings to neutral and charged scalar fields of the form $g'(\dual{\lambda}\lambda)\phi^{2}$ and $g''(\dual{\lambda}\lambda)\Phi^{\dag}\Phi$. Interactions with the scalar fields provide a natural portal for mass dimension one particles to interact with the SM sector through the Higgs boson. We note, as the Dirac field $\psi$ has mass dimension three-half, its quartic self-interaction would be of dimension six and is thus non-renormalizable in four space-time dimensions. The natural interaction between the Dirac and a neutral scalar field is the Yukawa interaction of the form $y\overline{\psi}\psi\phi$. Interactions of the form $y'\overline{\psi}\psi\phi^{2}$ and $y''\overline{\psi}\psi\Phi^{\dag}\Phi$ are of mass dimension five so they are non-renormalizable perturbatively.

The interactions that we have written down in the preceding paragraph involving the mass dimension one fields all share a common feature - They are all pseudo Hermitian. As a result, the Hermitian inner-product would not be invariant under time translation generated by these pseudo Hermitian Hamiltonians. This problem is not new and has been studied by Bender, Mostafazadeh, and collaborators~\cite{Bender:1998gh,Bender:1998ke,Bender:2002vv,Bender:2007nj,Mostafazadeh:2001jk,Mostafazadeh:2008pw}. There, the resolution is to introduce new inner-product that is invariant under time translation. For mass dimension one fields, defining the appropriate inner-product that preserves unitarity (in a generalized sense) is essential towards establishing the formalism to compute the $S$-matrix and physical observables. For now, this task remains incomplete and requires further investigations.

To the best of our knowledge, what we have presented here, for the first time, is a physical quantum field theory which realizes Wigner degeneracy with possibly important ramifications for physics beyond the SM.
While we have only studied the kinematics of the mass dimension one fields, resolving the problem of rotational covariance is an important milestone towards the ultimate goals of establishing consistent interacting theories and phenomenological models. It is our hope that one day, in the foreseeable future, we can make the definitive statement - \textit{Dark matter are described by the unitary irreducible representations of the extended Poincar\'{e} group with Wigner degeneracy.} This we believe, would be the best tributes we can give to Professor Eugene Wigner and Professor Steven Weinberg, who have contributed so much to our understanding of nature.

\appendix
\section{Reducible Hermitian formulation}\label{reducible}

With the introduction of Wigner degeneracy, the Elko spin sums, as computed with the Elko duals~(\ref{eq:dual1}-\ref{eq:dual2}), are Lorentz-invariant. Here, we explore what happens to Elko and the quantum fields, when we instead formulate the theory using the Dirac dual. As we will show, while the resulting quantum field theory is local and respects Lorentz symmetry, it furnishes a reducible representation of the inhomogeneous Lorentz group. Specifically, it is a direct sum of two irreducible representations, where each irreducible sector is described by the mass dimension one fields with Klein-Gordon kinematics.

Under the Dirac dual, we have
\begin{equation}
\overline{\xi}(\p,\tau)=\xi^{\dag}(\p,\tau)\gamma^{0},\quad
\overline{\zeta}(\p,\tau)=s\zeta^{\dag}(\p,\tau)\gamma^{0}.
\end{equation}
The spin sums are given by
\begin{align}
\sum_{\tau}\xi(\p,\tau)\overline{\xi}(\p,\tau)&=\gamma^{\mu}p_{\mu},\\
\sum_{\tau}\zeta(\p,\tau)\overline{\zeta}(\p,\tau)&=s\gamma^{\mu}p_{\mu}.
\end{align}
For now, we shall ignore the fact that the norms vanish with respect to the Dirac dual and proceed to construct a quantum field theory with $\lambda$ and $\overline{\lambda}=\lambda^{\dag}\gamma^{0}$. From~(\ref{eq:gp1}-\ref{eq:gp4}), we know that Elko do not satisfy the Dirac equation. We rewrite these identities as
\begin{equation}
\gamma^{\mu}p_{\mu}\xi(\p,\tau)=m\xi'(\p,\tau),\quad
\gamma^{\mu}p_{\mu}\zeta(\p,\tau)=-m\zeta'(\p,\tau).
\end{equation}
where
\begin{align}
\xi'(\p,1)=+i\xi(\p,2),\quad \zeta'(\p,1)=+i\zeta(\p,2),\\
\xi'(\p,2)=-i\xi(\p,1),\quad \zeta'(\p,2)=-i\zeta(\p,1),\\
\xi'(\p,3)=-i\xi(\p,4),\quad \zeta'(\p,3)=-i\zeta(\p,4),\\
\xi'(\p,4)=+i\xi(\p,3),\quad \zeta'(\p,4)=+i\zeta(\p,3).
\end{align}
Next, we define a new quantum field
\begin{equation}
\lambda'(x)=(2\pi)^{-3/2}\int\frac{d^{3}p}{\sqrt{2E}}\sum_{\tau}\left[e^{-ip\cdot x}\xi'(\p,\tau)a(\p,\tau)+e^{ip\cdot x}\zeta'(\p,\tau)b^{\dag}(\p,\tau)\right],
\end{equation}
so that the action of the Dirac operator on $\lambda$ yields
\begin{equation}
i\gamma^{\mu}\partial_{\mu}\lambda(x)=m\lambda'(x). \label{eq:D1}
\end{equation}

Repeat the above calculations with $\xi'$ and $\zeta'$, we find
\begin{align}
\sum_{\tau}\xi'(\p,\tau)\overline{\xi}'(\p,\tau)&=\gamma^{\mu}p_{\mu},\\
\sum_{\tau}\zeta'(\p,\tau)\overline{\zeta}'(\p,\tau)&=s\gamma^{\mu}p_{\mu},
\end{align}
and
\begin{equation}
i\gamma^{\mu}\partial_{\mu}\lambda'(x)=m\lambda(x).\label{eq:D2}
\end{equation}

The fields $\lambda$ and $\lambda'$ do not satisfy the Dirac equation. But by combining them, we can introduce a set of eight-component fields, namely
\begin{equation}
\Lambda(x)=\left[\begin{matrix}
    \lambda(x) \\
    \lambda'(x)
\end{matrix}\right],\quad
\overline{\Lambda}(x)=
\left[\begin{matrix}
    \overline{\lambda}'(x) & \overline{\lambda}(x)
\end{matrix}\right],
\end{equation}
such that $\Lambda$ satisfies the Dirac equation
\begin{equation}
\left(i\Gamma^{\mu}\partial_{\mu}-M\right)\Lambda(x)=\mathbb{O},
\end{equation}
where
\begin{equation}
    \Gamma^{\mu}=\left(\begin{matrix}
        \mathbb{O} & \gamma^{\mu} \\
        \gamma^{\mu} & \mathbb{O}
    \end{matrix}\right),\quad
    M=\left(\begin{matrix}
        m\mathbb{I} & \mathbb{O} \\
        \mathbb{O} & m\mathbb{I}
    \end{matrix}\right).
\end{equation}
Therefore, we obtain the following Lagrangian density
\begin{equation}
\mathscr{L}_{\Lambda}=\overline{\Lambda}\left(i\Gamma^{\mu}\partial_{\mu}-M\right)\Lambda. \label{eq:L_Lambda}
\end{equation}
Using~(\ref{eq:L_Lambda}), we can readily show that the fields and their conjugate momenta satisfy the canonical commutation/anti-commutation relations and that the free Hamiltonian is positive-definite.

\subsubsection*{Decomposing the Lagrangian density}

The fields $\Lambda$ and $\overline{\Lambda}$ seem to present a new quantum field theory that are physically distinct to $\lambda$ and $\dual{\lambda}$ owing to the difference in kinematics and mass dimensions. However, these differences turn out to be superficial. Here, we show that the Lagrangian density $\Lambda$ and $\overline{\Lambda}$ decomposes into a sum of two physically equivalent Lagrangian densities of mass dimension one fields. Therefore, the field theory as described by $\mathscr{L}_{\Lambda}$ is reducible.

To prove our statement, let us write the free Lagrangian density of $\dual{\lambda}$ and $\lambda$ as
\begin{equation}
\mathscr{L}(\dual{\lambda},\lambda)=\left(\partial^{\mu}\dual{\lambda}\partial_{\mu}\lambda-m^{2}\dual{\lambda}\lambda\right).\label{eq:L}
\end{equation}
Because of the $\dual{\lambda}\lambda$ product, the Lagrangian density is pseudo Hermitian. Despite the pseudo-Hermiticity, as we have shown in~\ref{qf}, the theory is local, respects Lorentz symmetry, and the free Hamiltonian is Hermitian and positive-definite. These features, along with the results given in sec.~\ref{discrete} tell us that~\ref{eq:L} is the correct irreducible Lagrangian density for mass dimension one fields. By irreducible, we mean that $\mathscr{L}(\dual{\lambda},\lambda)$ cannot be decomposed into a sum of Lagrangian densities consisting of other independent fields.

We will now demonstrate that $\mathscr{L}_{\Lambda}$ is reducible. For this purpose, we first use the identity  $\dual{\lambda}=\overline{\lambda}'$ to rewrite $\mathscr{L}(\dual{\lambda},\lambda)$ as
\begin{align}
\mathscr{L}(\dual{\lambda},\lambda)=\mathscr{L}(\overline{\lambda}',\lambda) 
=\partial^{\mu}\overline{\lambda}'\partial_{\mu}\lambda-m^{2}\overline{\lambda}'\lambda
\end{align}
Next, we add a Hermitian conjugate to $\mathscr{L}(\overline{\lambda}',\lambda)$ to obtain a Hermitian Lagrangian density
\begin{align}
\mathscr{L}_{h}&=\mathscr{L}(\overline{\lambda}',\lambda)+\mathscr{L}^{\dag}(\overline{\lambda}',\lambda) \nonumber\\
&=\left(\partial^{\mu}\overline{\lambda}'\partial_{\mu}\lambda-m^{2}\overline{\lambda}'\lambda\right)+
\left(\partial^{\mu}\overline{\lambda}\partial_{\mu}\lambda'-m^{2}\overline{\lambda}\lambda'\right).\label{eq:L_H}
\end{align}
As the first and second term in~(\ref{eq:L_H}) are physically equivalent, $\mathscr{L}_{h}$ is reducible. That is, for the purpose of studying mass dimension one fields, $\mathscr{L}(\overline{\lambda}',\lambda)$ suffices. The reason for constructing a reducible $\mathscr{L}_{h}$ is to show that it is equal to $m\mathscr{L}_{\Lambda}$ up to total derivative terms. Integrating $\mathscr{L}_{h}$ by parts and ignore the total derivative terms, we obtain
\begin{equation}
\mathscr{L}_{h}=-\overline{\lambda}'(\partial^{\mu}\partial_{\mu}+m^{2})\lambda
-\overline{\lambda}(\partial^{\mu}\partial_{\mu}+m^{2})\lambda'.\label{eq:L_H2}
\end{equation}
Using the identity
\begin{equation}
    (\partial^{\mu}\partial_{\mu}+m^{2})\mathbb{I}=(i\gamma^{\mu}\partial_{\mu}+m\mathbb{I})(-i\gamma^{\nu}\partial_{\nu}+m\mathbb{I}),
\end{equation}
the first term in~(\ref{eq:L_H2}) can be written as
\begin{align}
-\overline{\lambda}'(\partial^{\mu}\partial_{\mu}+m^{2})\lambda
&=-\overline{\lambda}'(i\gamma^{\mu}\partial_{\mu}+m)(-i\gamma^{\nu}\partial_{\nu}+m)\lambda \nonumber\\
&=-\overline{\lambda}'(i\gamma^{\mu}\partial_{\mu}+m)(-m\lambda'+m\lambda)\nonumber\\
&=	m\overline{\lambda}'(i\gamma^{\mu}\partial_{\mu}+m)\lambda'-m\overline{\lambda}'(i\gamma^{\mu}\partial_{\mu}+m)\lambda\nonumber\\
&=m\overline{\lambda}'(i\gamma^{\mu}\partial_{\mu}+m)\lambda'-m^{2}(\overline{\lambda}'\lambda'+\overline{\lambda}'\lambda)\nonumber\\
&=m\left(\overline{\lambda}'i\gamma^{\mu}\partial_{\mu}\lambda'-m\overline{\lambda}'\lambda\right).
\end{align}
Repeating the same calculations for the second term of $\mathscr{L}_{h}$, we find
\begin{align}
\mathscr{L}_{h}&=m\left[\left(\overline{\lambda}'i\gamma^{\mu}\partial_{\mu}\lambda'-m\overline{\lambda}'\lambda\right)
+\left(\overline{\lambda}i\gamma^{\mu}\partial_{\mu}\lambda-m\overline{\lambda}\lambda'\right)\right] \nonumber\\
&=m\mathscr{L}_{\Lambda}.
\end{align}
Since $\mathscr{L}_{h}$ is reducible, it follows that $\mathscr{L}_{\Lambda}$ is also reducible.

\section{The solution for $\pmb{\eta}$}\label{eta}

To solve $\eta$, let us start with
\begin{equation}
\eta_{a}=e^{i\theta_{12} N_{12}}e^{i\theta_{34} N_{34}},
\end{equation}
where
\begin{align}
N_{12}&=\int d^{3}p\left[a^{\dag}(\p,2)a(\p,1)-a^{\dag}(\p,1)a(\p,2)\right],\\
N_{34}&=\int d^{3}p\left[a^{\dag}(\p,4)a(\p,3)-a^{\dag}(\p,3)a(\p,4)\right].
\end{align}
Acting $\eta_{a}$ on the state $|p,\tau,a\rangle$, we get
\begin{align}
\eta_{a}|p,1,a\rangle&=\cos\theta_{12}|p,1,a\rangle+i\sin\theta_{12}|p,2,a\rangle,\label{eq:eta_a1}\\
\eta_{a}|p,2,a\rangle&=\cos\theta_{12}|p,2,a\rangle-i\sin\theta_{12}|p,1,a\rangle,\\
\eta_{a}|p,3,a\rangle&=\cos\theta_{34}|p,3,a\rangle+i\sin\theta_{12}|p,4,a\rangle,\\
\eta_{a}|p,4,a\rangle&=\cos\theta_{34}|p,4,a\rangle-i\sin\theta_{12}|p,3,a\rangle.\label{eq:eta_a4}
\end{align}
Similarly, we take
\begin{equation}
\eta_{b}=e^{i\phi_{12} M_{12}}e^{i\phi_{34} M_{34}},
\end{equation}
where
\begin{align}
M_{12}&=\int d^{3}p\left[b^{\dag}(\p,2)b(\p,1)-b^{\dag}(\p,1)b(\p,2)\right],\\
M_{34}&=\int d^{3}p\left[b^{\dag}(\p,4)b(\p,3)-b^{\dag}(\p,3)b(\p,4)\right],
\end{align}
to obtain
\begin{align}
\eta_{b}|p,1,b\rangle&=\cos\phi_{12}|p,1,b\rangle+i\sin\phi_{12}|p,2,b\rangle,\label{eq:eta_b1}\\
\eta_{b}|p,2,b\rangle&=\cos\phi_{12}|p,2,b\rangle-i\sin\phi_{12}|p,1,b\rangle,\\
\eta_{b}|p,3,b\rangle&=\cos\phi_{34}|p,3,b\rangle+i\sin\phi_{34}|p,4,b\rangle,\\
\eta_{b}|p,4,b\rangle&=\cos\phi_{34}|p,4,b\rangle-i\sin\phi_{34}|p,3,b\rangle.\label{eq:eta_b4}
\end{align}
Define
\begin{equation}
\eta_{ab}=\eta_{a}\eta_{b}.
\end{equation}
Because $\eta_{a}|0\rangle=|0\rangle$, $\eta_{b}|0\rangle=|0\rangle$ and $[\eta_{a},\eta_{b}]=0$, the actions of $\eta_{ab}$ on the states are identical to~(\ref{eq:eta_a1}-\ref{eq:eta_a4}) and~(\ref{eq:eta_b1}-\ref{eq:eta_b4}). Next, we note
\begin{align}
\left(\eta_{ab}-\eta^{-1}_{ab}\right)|p,1,a\rangle&=+2i\sin\theta_{12}|p,2,a\rangle,\\
\left(\eta_{ab}-\eta^{-1}_{ab}\right)|p,2,a\rangle&=-2i\sin\theta_{12}|p,1,a\rangle,\\
\left(\eta_{ab}-\eta^{-1}_{ab}\right)|p,3,a\rangle&=+2i\sin\theta_{34}|p,4,a\rangle,\\
\left(\eta_{ab}-\eta^{-1}_{ab}\right)|p,4,a\rangle&=-2i\sin\theta_{34}|p,3,a\rangle,
\end{align}
and
\begin{align}
\left(\eta_{ab}-\eta^{-1}_{ab}\right)|p,1,b\rangle&=-2i\sin\phi_{12}|p,2,b\rangle,\\
\left(\eta_{ab}-\eta^{-1}_{ab}\right)|p,2,b\rangle&=+2i\sin\phi_{12}|p,1,b\rangle,\\
\left(\eta_{ab}-\eta^{-1}_{ab}\right)|p,3,b\rangle&=-2i\sin\phi_{34}|p,4,b\rangle,\\
\left(\eta_{ab}-\eta^{-1}_{ab}\right)|p,4,b\rangle&=+2i\sin\phi_{34}|p,3,b\rangle.
\end{align}
Therefore, the solution for $\eta$ is given by
\begin{equation}
\eta=\frac{1}{2}\left(\eta_{ab}-\eta^{-1}_{ab}\right)\label{eq:eta_soln}
\end{equation}
with
\begin{equation}
\theta_{12}=-\theta_{34}=\frac{\pi}{2},\quad
\phi_{12}=-\phi_{34}=\frac{\pi}{2},
\end{equation}
where
\begin{equation}
\eta_{ab}=\exp\left\{\frac{i\pi}{2}\int d^{3}p\sum_{\tau}\epsilon(\tau)\left[a^{\dag}(\p,\rho(\tau))a(\p,\tau)+b^{\dag}(\p,\rho(\tau))b(\p,\tau)\right]\right\},\label{eq:eta_ab}
\end{equation}
and
\begin{equation}
\epsilon(1)=-\epsilon(2)=\epsilon(3)=-\epsilon(4)=1.
\end{equation}

%{\color{red}{Is the $\eta$ operator is Lorentz invariant? To show this, we act the unitary Lorentz transformation operator $U(\Lambda)$ on $\eta_{ab}$ to obtain
%\begin{align}
%U(\Lambda)\eta_{ab}U^{-1}(\Lambda)&=
%\exp\left\{\frac{i\pi}{2}\int d^{3}p\sum_{\tau}\epsilon(\tau)U(\Lambda)\left[a^{\dag}(\p,\rho(\tau))a(\p,\tau)+b^{\dag}(\p,\rho(\tau))b(\p,\tau)\right]U^{-1}(\Lambda)\right\}.
%\end{align}
%The Lorentz transformation on the creation operators are~\cite{Weinberg:1995mt}
%\begin{align}
%U(\Lambda)a^{\dag}(\p,\tau)U^{-1}(\Lambda)&=\sqrt{\frac{E_{\Lambda p}}{E_{p}}}\sum_{\tau'}D_{\tau'\tau}(W(\Lambda,p))a^{\dag}(\boldsymbol{\Lambda p},\tau'),\\
%U(\Lambda)b^{\dag}(\p,\tau)U^{-1}(\Lambda)&=\sqrt{\frac{E_{\Lambda p}}{E_{p}}}\sum_{\tau'}D_{\tau'\tau}(W(\Lambda,p))b^{\dag}(\boldsymbol{\Lambda p},\tau').
%\end{align}
%where $W(\Lambda,p)$ is the element of the little group.}}

\acknowledgments

GBG thanks FAPESP (2021/12126-5) for financial support and Dharam Vir Ahluwalia for his great scientific legacy. He dedicates this work to the memory of Dharam Vir Ahluwalia. JMHS thanks CNPq, Brazil for financial support through grant number 307641/2022-8. CYL thanks James Brister, Chou-Man Sou, Zheng Sun, Wenqi Yu and Siyi Zhou for useful discussions. CYL is supported by The Sichuan University Post-doctoral Research Fund No.~2022SCU12119. BMP thanks CNPq for partial support.

\bibliography{Bibliography}
\bibliographystyle{JHEP}


\end{document}